\documentclass[a4paper,twocolumn,11pt, accepted=2022-08-25]{quantumarticle}
\pdfoutput=1
\usepackage[square,sort&compress,comma,numbers]{natbib}
\usepackage[utf8]{inputenc}
\usepackage[english]{babel}
\usepackage[T1]{fontenc}
\usepackage{hyperref}

\usepackage{amsmath}
\usepackage{amssymb}
\usepackage{bbm}
\usepackage{graphicx}
\usepackage{amsfonts}
\usepackage{color}
\usepackage{upgreek}
\usepackage{mathtools}
\usepackage{braket}
\makeatletter
\def\captionof#1#2{{\def\@captype{#1}#2}}
\makeatother
\begin{document}

\title{Periodically refreshed quantum thermal machines}

\author{Archak Purkayastha}
\email{archak.p@phys.au.dk}
\affiliation{School of Physics, Trinity College Dublin, College Green, Dublin 2, Ireland}
\affiliation{Centre for complex quantum systems, Aarhus University, Nordre Ringgade 1, 8000 Aarhus C, Denmark}

\author{Giacomo Guarnieri}
\email{giacomo.guarnieri@fu-berlin.de}
\affiliation{Dahlem Center for Complex Quantum Systems, Freie Universit at Berlin, 14195 Berlin, Germany}

\author{Steve Campbell}
\email{steve.campbell@ucd.ie}
\affiliation{School of Physics, University College Dublin, Belfield, Dublin 4, Ireland}
\affiliation{Centre for Quantum Engineering, Science, and Technology, University College Dublin, Belfield, Dublin 4, Ireland}

\author{Javier Prior}
\email{javier.prior@um.es}
\affiliation{Departamento de F\'isica, Universidad de Murcia, Murcia E-30071, Spain}

\author{John Goold}
\email{gooldj@tcd.ie}
\affiliation{School of Physics, Trinity College Dublin, College Green, Dublin 2, Ireland}


\begin{abstract}
We introduce unique class of cyclic quantum thermal machines (QTMs) which
can maximize their performance at the finite value of cycle duration $\tau$ where they are most irreversible. These QTMs are based on single-stroke thermodynamic cycles realized by the non-equilibrium steady state (NESS) of the so-called Periodically Refreshed Baths (PReB) process. We find that such QTMs can interpolate between standard collisional QTMs, which consider repeated interactions with single-site environments, and autonomous QTMs operated by simultaneous coupling to multiple macroscopic baths.  We discuss the physical realization of such processes and show that their implementation requires a finite number of copies of the baths. Interestingly, maximizing performance by operating in the most irreversible point as a function of $\tau$ comes at the cost of increasing the complexity of realizing such a regime, the latter quantified by the increase in the number of copies of baths required. We demonstrate this physics considering a simple example.  We also introduce an elegant description of the PReB process for Gaussian systems in terms of a discrete-time Lyapunov equation. Further, our analysis also reveals interesting connections with Zeno and anti-Zeno effects.
\end{abstract}

\maketitle

\section{Introduction}

In traditional cyclic heat engines, maximum efficiency is obtained in the limit of infinitely slow cycles ~\cite{Leff_1975,Leff_2018,Bhattacharjee_2021,Myers_2022}. In this regime, the process becomes reversible and the entropy production rate associated with it vanishes. However, in the same limit, the power also goes to zero. This constitutes a trade-off between the power extracted and the efficiency of the heat engine as a function of the cycle duration $\tau$. An analogous trade-off holds between the cooling rate and the coefficient of performance for traditional cyclic refrigerators. 

\begin{figure*}
\centering
\includegraphics[width=0.98\textwidth]{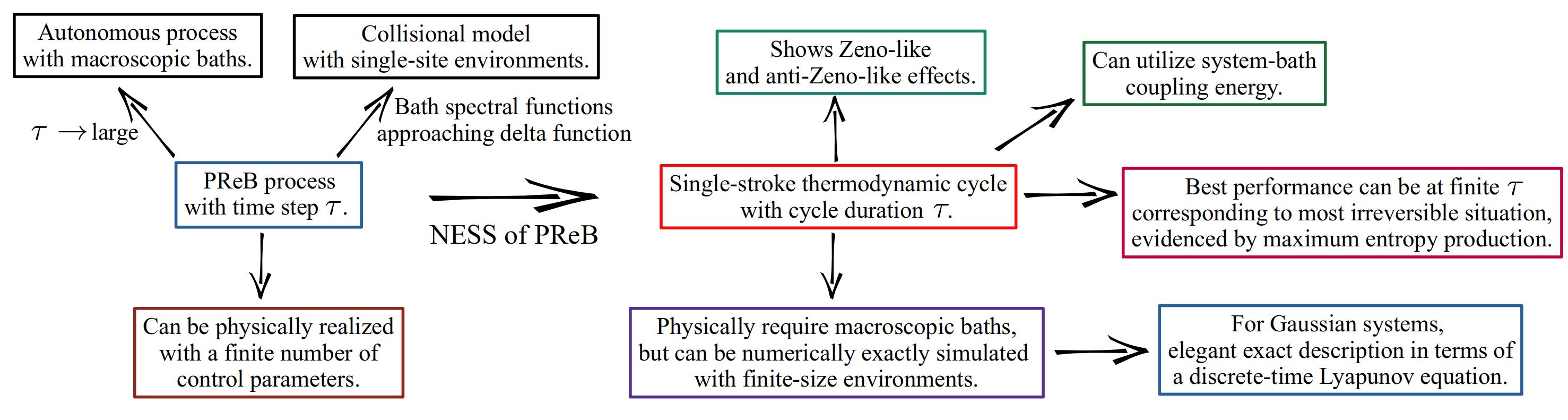}
\caption{\label{fig:flowchart} The summary of our main results. Periodically refreshed QTMs are based on the single-stroke thermodynamic cycle realized at the NESS of PReB. The `best performance' in above diagram refers to simultaneous maximization of efficiency and power output in heat engine regime, simultaneous maximization of coefficient of performance and cooling rate in refrigerator regime. }
\end{figure*}

In this paper, we introduce a different type of single-stroke thermodynamic cycle for which, in the heat engine regime, efficiency can be maximized at the finite value of $\tau$ which corresponds to the most irreversible situation, evidenced by a peak in the entropy production rate. As a result, efficiency and extracted power of these heat engines can be simultaneously maximized without any trade-off as a function of $\tau$. Analogous statements hold for the refrigerating regime. Thus, quantum thermal machines (QTMs) based on this unique type of thermodynamic cycle can harness irreversibility to significantly boost their performance. 

This unique type of thermodynamic cycle corresponds to the non-equilibrium steady state (NESS) of the so-called Periodically Refreshed Baths (PReB) process \cite{PReB}. The PReB process describes a framework where a system is coupled to multiple macroscopic thermal baths which, from the perspective of the system, are refreshed to their original thermal states in steps of time $\tau$. This differs from a continuous-time autonomous process where the baths are always connected and never refreshed, which is the standard scenario in open quantum systems \cite{Breuer_book,Jauho_book,Kamenev_book}. We discuss how the PReB process can be physically realized in practice, and interestingly, discover a different type of trade-off.  We find that boosting performance by going to the most irreversible point as a function of $\tau$, comes at the cost of an increase in complexity of physically realizing such regimes.

The mathematical structure of the PReB process puts it in the class of collisional or repeated interaction models \cite{RauPR,Campbell_2021,Ciccarello_2021}. These models have emerged as versatile tools for fundamental understanding of open quantum systems and quantum thermodynamics (see, for example, \cite{Cattaneo_2021,Strasberg_PRX_2017,Strasberg_PRL_2019,Gabriele_NJP_2018,guarnieri2020non,
Barra_SciRep_2015}). Standard collisional models provide the dynamics of a system which repeatedly interacts with identically prepared single-site environments. Including macroscopic baths described by spectral functions into this approach has remained a challenge. The PReB process offers an efficient way to do so.  

We find that when the bath spectral function approaches a Dirac-delta function, the PReB process reduces to standard collisional models with single-site environments. Meanwhile, it has been shown that for increasing $\tau$, the state obtained from the PReB process converges to that of the continuous-time autonomous process \cite{PReB}. Thus, the unique QTMs at the NESS of the PReB process can interpolate between QTMs based on standard collisional models and autonomous QTMs realized at the NESS in the presence of multiple macroscopic baths (for example, absorption refrigerators~\cite{Mark_2019} and thermoelectric devices~\cite{Benenti_2017}).

To explore the unique properties of QTMs realized at the NESS of the PReB process, we require an efficient way to find NESS of the PReB process. We introduce a powerful and elegant way to describe the PReB process (and hence for any collisional or repeated interaction process) for Gaussian systems in terms of a discrete time Lyapunov equation. The discrete time Lyapunov equation is well-studied in mathematics and engineering, and efficient methods for solving it are available \cite{Control_theory_book1,Control_theory_book2}. The description in terms of the discrete time Lyapunov equation  allows us to directly obtain the exact NESS of the PReB process without actually carrying out the full time evolution. This description holds for arbitrary Gaussian systems in a lattice of arbitrary dimension and geometry. We apply it to a simple example to demonstrate the physics discussed above.

It is important to mention that we consider no restriction on the strength of system-bath coupling. Rather, the unique cyclic QTMs at the NESS of PReB utilize the switching on and off of the system-bath couplings as a source of power. In fact, in the simple example we use to explore both heat engine and refrigerator regimes of operation, this is the only source of power. In many works, changes in system-bath coupling energy are not considered,  either due to using a weak system-bath coupling approximation, or by explicitly imposing conservation of the system-bath coupling Hamiltonian \cite{Dann_2021_Quantum, Dann_2021_PRR}.  Various other works have explored the effect of time-dependent system-bath couplings on the ensuing thermodynamics (for example, \cite{Esposito_2015_1,Esposito_2015_2,Newman_2017,Restrepo_2018,Marti_2018, Schaller_2018,Pancotti_2020}). However, in most of these cases the external driving of system-bath couplings is not the sole source of power. The use of system-bath coupling as the sole source of power to perform thermodynamic tasks distinguishes our example from all such cases. Utilizing driven system-bath coupling energy as the only source of power to perform thermodynamic tasks has been recently explored in a completely different setting \cite{Carrega_2022}.

There is also no restriction on the magnitude of $\tau$ and thus, the periodic switching on and off of the system-bath couplings can be arbitrarily fast or slow. In fact, using the simple example, we study the properties of the QTMs as function of $\tau$. We find that when $\tau$ is smaller than system time-scales, a Zeno-like effect \cite{Misra_Sudarshan_1977} happens where the dynamics is slowed down with decreasing $\tau$. In the Zeno regime, the efficiency saturates while the power goes to zero. The simultaneous enhancement of efficiency and power is seen in the anti-Zeno-like regime \cite{Kofman_2000,Kofman_2001,Kofman_2004,Erez_2008,Alvarez_2010,Rao_2011,
Mukherjee_2020,Das_2020}, where increasing the rate of periodic refreshing speeds up the dynamics. 

A summary of main results is given in Fig.~\ref{fig:flowchart}.
Our results enhance fundamental understanding of open quantum systems and quantum thermodynamics in a wide range of directions, and connect various seemingly disparate concepts. Moreover, our proposal for physically realizing the PReB process requires a finite number of control parameters, making the experimental implementation of the unique thermodynamic cycle plausible. This may in turn open different pathways for the efficient management of energy at microscopic levels, which is crucial for the development of quantum technology~\cite{Affeves_2021}. 

The paper is arranged as follows. First, in Sec.~\ref{Sec:PReB}, we recap the PReB process. Then, in Sec.~\ref{Sec: Physically_engineering_PReB} we discuss the physical realization of the PReB process. In Sec.~\ref{Sec:NESS_of_PReB_as_cyclic_process}, we discuss how a unique type of thermodynamic cycle is realized at the NESS of the PReB process. In Sec.~\ref{Sec:the example}, we introduce the simple example that we use in the paper to demonstrate the counter-intuitive features of the thermodynamic cycle. In order to do so, in Sec.~\ref{Sec:Formalism_for_thermodynamics}, we give the definitions of thermodynamic quantities at NESS. In Sec.~\ref{Sec:Lyapunov}, we show how, for Gaussian systems, the thermodynamic quantities at NESS can be obtained exactly via solving a discrete-time Lyapunov equation. In Sec.~\ref{Sec:heat engine and refrigerator regimes}, we apply the formalism to our example, exploring the heat engine and the refrigerator regimes of operation and showing that the best performance in both regimes correspond to the most irreversible situation as a function of cycle duration. In Sec.~\ref{Sec: conditions for heat engine and refrigerator regimes}, by exploiting the connection with standard collisional models based on single-site environments, we give the necessary conditions for the heat engine and the refrigerator regimes of operation in our example. In Sec.~\ref{Sec: Zeno and anti-Zeno}, we discuss the connection with Zeno and anti-Zeno effects. In Sec.~\ref{Sec:Trade-off}, we discuss how boosting performance by operating in the most irreversible regime comes at the cost of an increase in complexity of realization. In Sec.~\ref{Sec: where is the heat extracted form}, we discuss the fundamental aspect of where the heat is extracted from in these QTMs. In Sec.~\ref{Sec:conclusions}, we summarize our results and provide some outlook. This is followed by several Appendices that provide additional technical details of various derivations.

\section{Recap of Periodically refreshed baths (PReB)}
\label{Sec:PReB}

\subsection{Definition of the process}
\label{SubSec:Defintion_of_PReB_process}
 We consider a microscopic quantum system connected to multiple macroscopic environments (baths). The set-up is described by a Hamiltonian of the form
\begin{align}
\hat{H}= \hat{H}_S + \sum_{\ell}\hat{H}_{SB_\ell} + \sum_{\ell}\hat{H}_{B_\ell},
\end{align}
where $\hat{H}_S$ is the Hamiltonian of the system, $\hat{H}_{B_\ell}$ is the Hamiltonian of the $\ell$th bath, and $\hat{H}_{SB_\ell}$ is the  Hamiltonian describing the coupling between the system and the $\ell$th bath. Initially the system is in an arbitrary state while the baths are in their own respective thermal states with their own temperatures and chemical potentials,
\begin{align}
\label{initial_state}
&\hat{\rho}_{\rm tot}(0) = \hat{\rho}(0) \hat{\rho}_B(0),\nonumber \\
&\hat{\rho}_B(0) = \prod_\ell \frac{e^{-\beta_\ell(\hat{H}_{B_\ell}-\mu_\ell \hat{N}_{B_\ell})}}{Z_{B_\ell}},
\end{align}
where $\hat{\rho}(0)$ is the initial state of the system, while $\hat{\rho}_B(0)$ is the initial composite state of the all the baths, $\beta_\ell$ is the inverse temperature of the $\ell$th bath, $\mu_\ell$ is the chemical potential of the $\ell$th bath, $\hat{N}_{B_\ell}$ is the total particle number operator of the $\ell$th bath, and $Z_{B_\ell}$ is the corresponding partition function. Let $D_B$ be the number of degrees of freedom, or the size of each bath. Since the baths are macroscopic, $D_B \rightarrow \infty$. The state of the system after a time $t$ is given by
\begin{align}
\label{def_Lambda}
\hat{\rho}(t) &= \hat{\Lambda}(t)[\hat{\rho}(0)] \nonumber \\
&= \lim_{D_B\rightarrow\infty}{\rm Tr}_B\left(e^{-i\hat{H}t} \hat{\rho}(0)\hat{\rho}_B(0) e^{i\hat{H}t} \right),
\end{align}
where ${\rm Tr}_B(\ldots)$ refers to trace over bath degrees of freedom. In the above $\hat{\Lambda}(t)[\ldots]$ is a completely positive trace preserving (CPTP) map that takes any system state, connects it to the baths prepared in state $\rho_B(0)$ and evolves it up to a time $t$. 

Having defined the CPTP map, the PReB process~\cite{PReB} is described by
\begin{align}
\label{PReB}
\hat{\rho}^{\rm PReB}(n\tau) =\underbrace{\hat{\Lambda}(\tau) [\ldots[\hat{\Lambda}(\tau)[}_{\text{n times}}[\hat{\rho}(0)]]]\ldots].
\end{align}
Physically this describes the following process. The system is connected to the baths and the entire set-up is evolved up to a time $\tau$. At time $\tau$, the baths are detached and refreshed to their original initial states. Then the set-up is evolved again for a time $\tau$. This detaching and refreshing is subsequently done in steps of time $\tau$. This physical understanding validates the name. 

The mathematical structure of repeated applications of the same CPTP map on a system makes the PReB process a class of collisional or repeated interaction models~\cite{Ciccarello_2021,Campbell_2021,Cattaneo_2021,Strasberg_PRX_2017,
Gabriele_NJP_2018}. However, traditional collisional models typically consider repeated interactions with identically prepared single-site environments. In contrast, the PReB process considers repeated interactions with macroscopic baths. This difference allows us to include information regarding the bath spectral functions in the PReB process, which will be crucial for our purpose. 

Furthermore, a large body of works regarding traditional collisional or repeated interaction models deals with obtaining a description in terms of Lindblad equations in the limit of small $\tau$ \cite{Ciccarello_2021,Campbell_2021,Cattaneo_2021}. In order to derive the Lindblad equation, the system-bath coupling Hamiltonian is also scaled by $\tau^{-1/2}$. In the PReB process this scaling of the system-bath coupling Hamiltonian is not done. Our regime of interest will also not be limited to small $\tau$.  We will keep $\tau$ a free parameter and study various effects as a function of $\tau$. 

It has been proven~\cite{PReB} that if there is a unique steady state of the continuous-time dynamics generated by $\hat{\Lambda}(t)$, the state of the system obtained by the PReB process will converge to that obtained by the continuous-time dynamics for increasing $\tau$.  More precisely, 
\begin{align}
\label{PReB_convergence}
&\left|\left|\hat{\rho}^{\rm PReB}(n\tau) - \hat{\rho}(n\tau)\right|\right|=\epsilon(\tau),\nonumber \\
&\textrm{$\epsilon(\tau)$ decreases with increase in $\tau$},
\end{align}
where $\left| \left| \hat{O} \right | \right |$ is the norm of $\hat{O}$.
This observation was used to develop a numerically efficient way to obtain the full non-Markovian dynamics of interacting quantum many-body systems~\cite{PReB}. Here, by decreasing $\tau$ below the system time-scales we will enter a regime where the system does not have enough time to react to the presence of the baths before the baths are refreshed. This will be a Zeno-like regime \cite{Misra_Sudarshan_1977}, where the dynamics is slowed down. The interesting regime is then those of intermediate $\tau$, which we show can be engineered to be advantageous for QTMs.

\subsubsection{Finite-size baths via chain mapping}
\label{SubSubSec:Finite_size_baths}

Taking $D_B\rightarrow \infty$ in the definition of $\hat{\Lambda}(\tau)$ in Eq.(\ref{def_Lambda}) means $\hat{\Lambda}(\tau)$ describes the time evolution in the presence of macroscopic baths. Interestingly, for the canonical description of thermal baths in terms of an infinite number of bosonic or fermionic modes, one can get the same $\hat{\Lambda}(\tau)$ considering the baths to be particular finite-sized chains of size $L_B$ approximately proportional to $\tau$~\cite{PReB}. 
The canonical model for thermal baths is given as follows,
\begin{align}
\label{non_interacting_baths}
&\hat{H}_{B_\ell} = \sum_{r=1}^\infty \Omega_{r \ell} \hat{B}^\dagger_{r \ell} \hat{B}_{r \ell}, \nonumber \\
&\hat{H}_{SB_\ell} = \sum_{r=1}^\infty \left( \kappa_{r \ell} \hat{S}_\ell^\dagger \hat{B}_{r \ell} + \kappa_{r \ell}^*  \hat{B}_{r \ell}^\dagger \hat{S}_\ell \right), 
\end{align}
where $\hat{B}_{r \ell}$ is the bosonic or fermionic annihilation operator for the $r$th mode of the $\ell$th bath, and $\hat{S}_\ell$ is the system operator coupling to the $\ell$th bath. The dynamics of the system under the influence of such baths is completely governed by the bath spectral functions $\mathfrak{J}_\ell(\omega)$, and the Bose or the Fermi distribution functions corresponding to the initial states of the baths $\mathfrak{n}_\ell(\omega)$. They are defined as
\begin{align}
\label{def_bath_spectral_fermi_distribution}
&\mathfrak{J}_\ell(\omega)=\sum_{r=1}^{\infty} \left| \kappa_{r \ell} \right|^2 \delta(\omega - \Omega_{r \ell}),\nonumber \\
&\mathfrak{n}_\ell(\omega) = \frac{1}{e^{\beta_\ell(\omega-\mu_\ell)}\mp 1},
\end{align} 
where the upper sign is for bosons and the lower sign is for fermions. Any given bath spectral function with finite high and low frequency cut-offs can be mapped onto a nearest neighbour tight-binding chain with the first site attached to the system \cite{Prior,rc_mapping_bosons,rc_mapping_old,rc_mapping_QTD_book,Chain_mapping_bosons_fermions,rc_mapping_fermions},
\begin{align}
\label{chain_mapping}
& \hat{H}_{B_\ell}= \sum_{p=1}^{L_B} \Big[\varepsilon_{p,\ell} \hat{b}_{p,\ell}^\dagger\hat{b}_{p,\ell} \nonumber \\
& \qquad \qquad  + g_{p,\ell}(\hat{b}_{p,\ell}^\dagger\hat{b}_{p+1,\ell}+\hat{b}_{p+1,\ell}^\dagger\hat{b}_{p,\ell})\Big],\nonumber \\
&\hat{H}_{SB_\ell}=g_{0,\ell}(\hat{b}_{1,\ell}^\dagger \hat{S}_\ell  + \hat{S}^\dagger_\ell \hat{b}_{1,\ell}).
\end{align}
With $L_B \rightarrow \infty$ the mapping is exact.
That is to say, one can choose the parameters $\varepsilon_{p,\ell}$ and $g_{p,\ell}$ such that, upon diagonalizing the single-particle Hamiltonian corresponding to this tight-binding chain, exactly Eq.~(\ref{non_interacting_baths}) is recovered.
The on-site potentials $\varepsilon_{p,\ell}$ and the hoppings $g_{p,\ell}$ are obtained from the following set of recursion relations
\begin{align}
\label{rc_map}
& g_{p, \ell}^2 = \frac{1}{2\pi} \int d\omega \mathfrak{J}_{p,\ell}(\omega),\nonumber \\
&\varepsilon_{p+1, \ell} = \frac{1}{2\pi g_{p, \ell}^2} \int d\omega~\omega \mathfrak{J}_{p,\ell}(\omega), \nonumber \\
&\mathfrak{J}_{p+1, \ell}(\omega)= \frac{4g_{p,\ell}^2 \mathfrak{J}_{p,\ell}(\omega) }{\left[\mathfrak{J}_{p,\ell}^H (\omega)\right]^2 +\left[\mathfrak{J}_{p,\ell}(\omega)\right]^2},~~
\end{align}
with $p$ going from $0$ to $L_B$, $\mathfrak{J}_{0,\ell}(\omega) =\mathfrak{J}_{\ell}(\omega) $ and $\mathfrak{J}_{p,\ell}^H (\omega)$ being the Hilbert transform of  $\mathfrak{J}_{p,\ell} (\omega)$,
\begin{align}
\mathfrak{J}_{p,\ell}^H(\omega)= \frac{1}{\pi}\mathcal{P}\int_{-\infty}^{\infty} d\omega^\prime \frac{\mathfrak{J}_{p,\ell}(\omega^\prime)}{\omega -\omega^\prime},
\end{align} 
where $\mathcal{P}$ denotes the principal value \cite{rc_mapping_QTD_book}. 
With finite high and low frequency cut-offs, the parameters $\varepsilon_{p,\ell}$ and $g_{p,\ell}$ quickly tend to constants for increasing $p$. Let these constants be $\varepsilon_{B}$ and $g_{B}$. These constants are proportional to the bandwidths of the baths. With a finite value of $L_B$, the remaining infinite number of modes are described by the residual spectral function, obtained from the above recursion relations.

After mapping to a chain, the map $\hat{\Lambda}(\tau)$ is obtained by switching on the system-bath coupling and evolving the full set-up for a time $\tau$ and tracing out the bath degrees of freedom. Due to Lieb-Robinson bounds \cite{Prior,Woods_2015,Woods_2016,Shiraishi_2017}, the information about this switching on of system-bath coupling spreads with a finite velocity proportional to $g_B$ across the tight-binding chains representing the baths. Thus, in a time $\tau$ only the first 
\begin{align}
\label{chain_size}
L_B \sim \lceil g_B \tau \rceil + L_0
\end{align}
 sites of the tight-binding chain are affected by the presence of the system, where $\lceil \ldots \rceil$ denotes the ceiling function, and $L_0$ is an integer independent of $\tau$.  Thus, the value of $\tau$ fixes the size of the tight-binding chain required to describe $\hat{\Lambda}(\tau)$ to a finite value. In other words, the  $\hat{\Lambda}(\tau)$ obtained from modelling baths as such finite-size chains is indistinguishable from the $\hat{\Lambda}(\tau)$ obtained with an infinite number of modes in the baths. 

Note that the continuous-time process described by $\hat{\Lambda}(t)$, Eq.~\eqref{def_Lambda}, for any finite time $t$ can also be modelled by a bath of finite size $L_B \sim g_B t$ exactly as above. However, increasing $t$ requires larger baths. Specifically, the limit $t\rightarrow \infty$ will require $L_B\rightarrow \infty$. It is important to mention that this chain mapping is a mathematical trick used for calculation and physical understanding. In general, the baths of the form in Eq.~(\ref{non_interacting_baths}) can be in any arbitrary geometry and dimension, as long they have continuous bath spectral functions.

\begin{figure*}
\includegraphics[width=\textwidth]{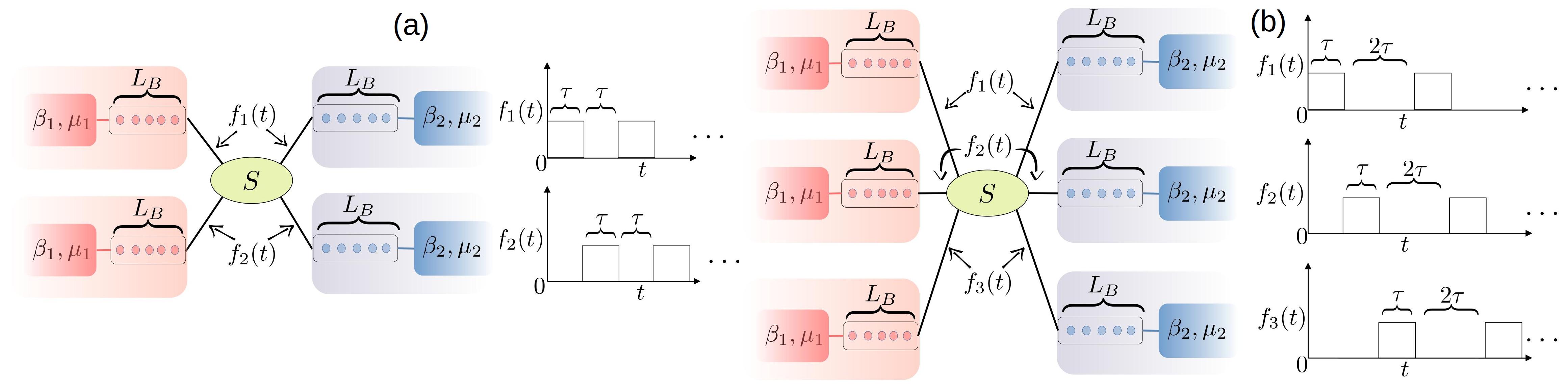}
\caption{\label{fig:schematic} {\bf Physically realizing the PReB process:} The figure gives a schematic for physically realizing the PReB process with two baths by engineering several copies of each bath. One copy of the baths are coupled with the system for a time $\tau$ which affects a finite part of each bath, represented by the finite-size chain of size $L_B$ (see Eq.(\ref{chain_size})) in the schematic. Then the coupling to these copies are switched-off and the coupling to the next copies are switched on. The first copies are then naturally re-thermalized in a time $\tau_R$ by the residual degrees of freedom left unaffected by the coupling with the system. {\bf (a)} Shows the case where  $\tau_R/\tau < 1$. In this case minimum two copies of each bath are required. The functions $f_1(t)$ and $f_2(t)$, which give the time dependence of system-bath couplings are shown in the plots. {\bf (b)} Shows the case when   $1<\tau_R/\tau < 2$. This requires engineering a minimum of three copies of each bath. The functions $f_1(t)$, $f_2(t)$ and $f_3(t)$ which give the time dependence of system-bath couplings are shown in the plots.  }
\end{figure*}

\section{Physically engineering the PReB process}
\label{Sec: Physically_engineering_PReB}
We can use the chain-mapping of bath spectral functions (Sec~\ref{SubSubSec:Finite_size_baths}) to understand how the PReB process may be physically realized. After carrying out the mapping onto $L_B$ sites, there is a residual spectral function, which describes the residual bath connected to the $L_B^{th}$ site. This residual bath has remained unaffected by the presence of the system up to a time $\tau$. If coupling to the system is switched-off at time $\tau$, the action of the residual bath will be to thermalize the chain of $L_B$ sites to its initial thermal state.  This crucially requires that the residual bath has infinite degrees of freedom while the chain is finite. In other words, since only a finite part of the macroscopic bath is affected in a time $\tau$, when disconnected from the system the remaining infinite degrees of freedom will rethermalize the part of the bath that has been affected by the system. Although the chain-mapping helps in making this understanding transparent, this will hold more generally, even for generic baths having many-body interactions.

Let $\tau_R$ be the time taken for such effective self-rethermalization of the bath. If ${\tau_R}/{\tau}<1$, then the PReB process with two baths at different temperatures and chemical potentials can be realized using a set-up of the form shown in Fig.~\ref{fig:schematic}(a). The set-up requires a minimum of two copies of each bath. The couplings between the copies of the baths and the system are periodically driven by specific square pulses, as shown. The coupling between one copy of each bath and the system is switched on for a time $\tau$. At time $\tau$, the couplings to these copies of the baths are switched-off and the coupling to the next copies are switched-on. While the system is being affected by the fresh copies during the next interval of time $\tau$, the first copies of the baths are refreshed by their corresponding residual baths. Thus, in steps of time $\tau$, the system sees a refreshed copy of the baths, thereby realizing the PReB process given in Eq.~(\ref{PReB}).

If ${\tau_R}/{\tau}>1$, realizing the PReB process requires engineering more than two copies of each bath. In general, the minimum number of copies $\mathfrak{N}$ of the baths required to realize a PReB process is given by
\begin{align}
\label{no_of_copies}
\mathfrak{N}=\left\lceil \frac{\tau_R}{\tau} \right\rceil +1.
\end{align}
The periodic driving of the coupling between the baths and the system then consists of a constant pulse for time $\tau$, which is then switched-off for a time $\left\lceil {\tau_R}/{\tau} \right\rceil$, during which the baths self-rethermalize to their original state. An example for the case of $\left\lceil {\tau_R}/{\tau} \right\rceil=2$ is shown in Fig.~\ref{fig:schematic}(b). 

Note that, during the entire process, the residual baths do not affect the dynamics of the system. Their only effect is to refresh the finite part of the bath that has been affected by the system. Hence, one can neglect the presence of the residual baths and consider only finite-size chains in describing the dynamics, although their presence is crucial to physical realization of the PReB process. Furthermore, the self-rethermalization of baths will be generically true even beyond the approximation of baths governed by quadratic Hamiltonians. In fact one does not need to engineer the microscopic details of the baths to realize the PReB process provided some estimate of the self-rethermalization time, $\tau_R$, is available. However, if specific bath spectral functions are desired, they can be engineered by designing a finite part of the bath. The parameters for such a design can be obtained via chain-mapping.  

It is intuitive that $\mathfrak{N}$ is proportional to the number of parameters that need to be precisely controlled for realization of the PReB process. The number $\mathfrak{N}$ then provides an ad-hoc notion of the complexity of realizing the PReB process. This understanding will be important in our discussion of QTMs at the NESS of the PReB process.

\section{NESS of PReB as a thermodynamic cycle}
\label{Sec:NESS_of_PReB_as_cyclic_process}

In this work we will assume that there is a unique steady state of the PReB process. Mathematically, this means that one eigenvalue of the CPTP map $\hat{\Lambda}(\tau)$ is $1$, while all other eigenvalues have magnitude less than $1$. Under such a condition, the PReB process, which is obtained by repeated action of the map, will eventually project the system to a state proportional to the eigenvector of $\hat{\Lambda}(\tau)$ corresponding to the eigenvalue $1$. This will be the NESS of the process, obtained in the limit of $n\rightarrow \infty$. The NESS density matrix satisfies,
\begin{align}
\label{def_PReB_NESS}
\hat{\rho}_{NESS}^{\rm PReB} = \hat{\Lambda}(\tau)\left[\hat{\rho}_{NESS}^{\rm PReB}\right].
\end{align}
Thus, the NESS of the PReB process corresponds to the state that is invariant under the action of $\hat{\Lambda}(\tau)$. This physically describes the following process. The system in state $\hat{\rho}_{NESS}^{\rm PReB}$ is connected to the thermal baths in the composite state $\hat{\rho}_B(0)$, the whole set-up is evolved for a time $\tau$, and then the system is detached and connected to refreshed copies of the baths. The final state of the set-up is exactly same as the initial state, therefore describing a cyclic process. During this cyclic process energy and particles have been exchanged with the thermal baths, thereby realizing a unique single-stroke thermodynamic cycle. This is fundamentally different from typical thermodynamic cycles, where at the end of every cycle the system is in equilibrium. In contrast, here, at the end of each cycle (and also during the cycle) the system is always out-of-equilibrium. If we can extract work out of this thermodynamic cycle, we will have a unique type of cyclic heat engine. If the thermodynamic cycle process can be used to cool a cold bath, we have a unique type of cyclic refrigerator.  It is interesting to note that, the cycle duration $\tau$ does not correspond to the time-period of the drive for the system-bath coupling required to realize the PReB process (see Fig.~\ref{fig:schematic}). 

\begin{figure*}
\centering
\includegraphics[width=\textwidth]{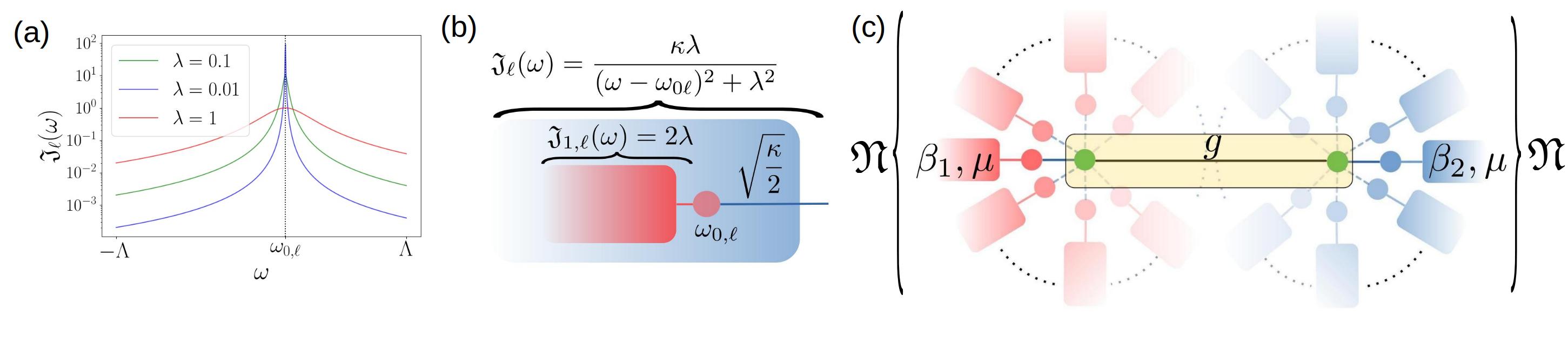}
\caption{\label{fig:two_site_example_schematic}{\bf The example set-up:} {\bf (a)} The chosen Lorentizian spectral functions, Eq.~\eqref{Lorentzian}, of the baths for various values of $\lambda$. {\bf (b)} The Lorentizian spectral functions can be thought of as arising from a single-site coupled to a residual bath with constant spectral function. This can be obtained via carrying out a single step of the chain-mapping procedure. {\bf (c)} The example we consider is two coupled fermionic sites, each with its own set of baths of the form in (b) which couple sequentially, i.e. one at a time, for a duration $\tau$, such that a PReB process is realized. The number of baths required to realize a PReB process is $\mathfrak{N}$. It can be argued that $\mathfrak{N}\gg [2\lambda \tau]^{-1}$. The baths attached to first (second) site has inverse temperature $\beta_1$ ($\beta_2$) while their chemical potentials, $\mu$ are the same.  }
\end{figure*}

From Eq.~(\ref{PReB_convergence}) we see that
\begin{align}
&\hat{\rho}_{NESS}^{\rm PReB} \rightarrow \lim_{t\rightarrow \infty} \left( \hat{\Lambda}(t)[\hat{\rho}(0)]\right)\textrm{ with increase in $\tau$}. 
\end{align}
Thus, on increasing $\tau$, the NESS of the PReB process will converge to the NESS of the continuous-time autonomous process. In fact, noting that the time evolution with baths modelled as chains of size $L_B\sim g_B \tau$ (Eq.(\ref{chain_size})) is indistinguishable from the time evolution with baths modelled as semi-infinite chains up to time $\tau$, we see that the NESS of the continuous-time autonomous process is obtained by taking $n=1$ and $\tau\rightarrow \infty$. Thus, it corresponds the infinitely slow limit of the thermodynamic cycle described by the NESS of PReB. Traditional cyclic QTMs become most efficient in this limit. In constrast, as we will demonstrate with an explicit example, the cyclic QTMs operating at the NESS of PReB can achieve their maximal efficiency for a {\it finite} value of $\tau$.

\section{The simple and insightful example}
\label{Sec:the example}

With the physical picture of how the PReB process can be realized in practice, we now introduce the example that we will mainly consider. This example will highlight various special features of the unique type of QTMs realized at the NESS of the PReB process.

\subsection{The system and the bath spectral functions}
As a simple example, we consider a system consisting of two fermionic sites with hopping between them, each site being attached to its own fermionic bath,
\begin{align}
&\hat{H}_S = g\left(\hat{c}_1^\dagger \hat{c}_2 +\hat{c}_2^\dagger \hat{c}_1 \right),  \\
&\hat{H}_{SB_\ell} = \sum_{r=1}^\infty \left( \kappa_{r \ell} \hat{c}_\ell^\dagger \hat{B}_{r \ell} + \kappa_{r \ell}^*  \hat{B}_{r \ell}^\dagger \hat{c}_\ell \right),~~\ell=\{1,2\}. \nonumber
\end{align}
The baths are of exactly the same form as Eq.~(\ref{non_interacting_baths}). They are periodically refreshed in steps of time $\tau$. This requires $\mathfrak{N}$ copies of each bath, see Fig.~\ref{fig:two_site_example_schematic}. We choose the bath spectral functions to be of the form
\begin{align}
\label{Lorentzian}
\mathfrak{J}_\ell(\omega) = 
\begin{cases}
\frac{\kappa \lambda}{(\omega-\omega_{0\ell})^2 +\lambda^2}~~\forall~~ &-\Lambda\leq\omega\leq\Lambda, \\
~~~~~~0&{\rm otherwise}.
\end{cases}
\end{align}
The bath spectral functions are therefore Lorentzian functions with a peak at $\omega_{0 \ell}$ and a width of $\lambda$, along with a hard cut-off at $\Lambda$. This serves as a simple minimal model for a structured bath spectral function. For small $\lambda$, the peak around $\omega_{0 \ell}$ leads to a highly non-Markovian dynamics of the system. Conversely, for $\lambda \gg \Lambda$ and $\kappa=\Gamma \lambda$, the baths become featureless, represented by a constant spectral function $\Gamma$ with hard cut-offs at $\pm \Lambda$. Increasing the value of $\lambda$ therefore allows us to go from a structured to a featureless bath (see Fig.~\ref{fig:two_site_example_schematic}(a)).

Further insight about the bath spectral function is provided by carrying out one step of the chain-mapping procedure, Eq.~(\ref{rc_map}).
For $\lambda \ll \Lambda$, one step of the chain-mapping procedure gives 
\begin{align}
\label{rc_mapping_Lorentzian}
g_{0,\ell} \approx \sqrt{\frac{\kappa}{2}},~~\varepsilon_{1,\ell} = \omega_{0\ell},~~\mathfrak{J}_{1,\ell}(\omega)\approx 2\lambda.
\end{align}
Thus, in this regime, each bath is effectively a single fermionic site, with on-site energy given by $\omega_{0\ell}$, that is coupled to a residual bath whose spectral function is a constant, $2\lambda$, see Fig.~\ref{fig:two_site_example_schematic}(b). The time taken to refresh and the number of copies of finite-size chains required to realize the PReB process, Eq.~\eqref{no_of_copies}, can therefore be estimated as
\begin{align}
\label{tau_R_Lorentizian}
\tau_R \gg \frac{1}{2\lambda},~~\mathfrak{N} \gg \frac{1}{2\lambda \tau}.
\end{align}
The whole set-up can be schematically represented as in Fig.~\ref{fig:two_site_example_schematic}(c).

For $\lambda \rightarrow 0$, i.e, when the bath spectral functions approach a delta function, Eqs.~(\ref{rc_mapping_Lorentzian}) and (\ref{tau_R_Lorentizian}) show that, each bath will effectively be only a single fermionic site, and an infinite number of copies will be needed. This is exactly what would be obtained in a traditional collisional or repeated interaction model. Convergence to the continuous-time NESS results are only possible for $\tau \gg (2\lambda)^{-1}$ such that the effect of the full spectral function, and not just that of a single-site, is observed.  The choice of bath spectral function in Eq.~(\ref{Lorentzian}) thus allows us to interpolate between the standard collisional or repeated interaction model limit with single-site environments and the continuous-time NESS with infinite baths, by changing the values of $\lambda$ and $\tau$. Studying the thermodynamics at the NESS of the PReB process for various values of $\lambda$ and $\tau$ is therefore insightful.

\subsection{Temperatures and chemical potentials}
We will assume that the bath attached to site $1$ is hot and concentrate on the case where there is no chemical potential difference between the two baths, i.e, 
\begin{align}
\label{temperature_and_chemical_potential}
\beta_1<\beta_2,~~\mu_1=\mu_2=\mu.
\end{align}
As we will discuss in more detail in the following, since there is no chemical potential bias, the only source of power at the NESS comes from the repeated external switching on and off of the system-bath couplings. At a finite value of $\tau$, this power can be used to realize both the heat engine and refrigerating regimes of operation for the thermodynamic cycle described by the NESS of the PReB process.
Notably, as we will see, these regimes of operation depend strongly on the nature of the bath spectral functions which are controlled by the value of $\lambda$.

In order to discuss the QTMs realized at the NESS, we first discuss the thermodynamics at NESS in general in the next section.

\section{Thermodynamics at the NESS}
\label{Sec:Formalism_for_thermodynamics}
 
\subsection{NESS of PReB process: $n\rightarrow \infty$, $\tau\rightarrow$ finite}
The NESS of the PReB process is obtained in the limit of $n\rightarrow \infty$ for a finite value of $\tau$. The NESS satisfies Eq.~(\ref{def_PReB_NESS}). Consequently, at the NESS the change in internal energy and the change in entropy of the system in a single step are zero. However, the work done on the system and the heat dissipated into the baths in a single step are non-zero constants. Dividing these single-step quantities by the cycle duration, $\tau$, then gives the cycle-averaged power input into the system and the heat currents into the baths, respectively.
In terms of these, the first law of thermodynamics at the NESS is given by
\begin{align}
\label{first_law_NESS}
& P=\sum_\ell \dot{Q}_{\ell},~~P = P_{\rm ext}+P_{\rm chem}
\end{align}
while the second law is given by
\begin{align}
\label{second_law_NESS}
\sigma=\sum_{\ell} \beta_\ell \dot{Q}_{\ell} \geq 0. 
\end{align}
Here, $P_{\rm ext}$ is the cycle-averaged power input into the system due to external switching on and off of the system-bath couplings at the NESS, $P_{\rm chem}$ is the cycle-averaged chemical power, $\dot{Q}_{\ell}$ is the cycle-averaged heat dissipation rate at the NESS, $\sigma$ is the cycle-averaged entropy production rate at the NESS. The sign convention for describing the thermodynamic behavior is given in Table~\ref{sign_convention}. 

\begin{table}
\begin{tabular}{cc}
$P>0$ & Power input \\ 
$P<0$ & Power extracted \\
$\dot{Q}_\ell>0$ & heat dissipated into $\ell$th bath \\
$\dot{Q}_\ell <0$ & heat extracted from $\ell$th bath 
\end{tabular}
\caption{\label{sign_convention} Our sign convention for describing thermodynamics.}
\end{table}

To obtain these quantities for the NESS of the PReB process, we define $\hat{H}_m$ as the Hamiltonian governing the dynamics of the system during the $m$th step,
\begin{align}
\label{finite_size_Hamiltonian}
\hat{H}_m = \hat{H}_S + \sum_{\ell}  \hat{H}_{B_{\ell m}} + \sum_\ell \hat{H}_{SB_{\ell m}},
\end{align}
where $\hat{H}_{B_{\ell m}}$ is the Hamiltonian of the finite-size chain of the $\ell$th bath that is connected to the system during the $m$th step, obtained via chain-mapping (see Eqs.~(\ref{chain_mapping}) and (\ref{chain_size})) and  $\hat{H}_{SB_{\ell m}}$ is the corresponding system-bath coupling. At the NESS of PReB, we have
\begin{align}
& \hat{\rho}_{\rm tot}^{NESS}(\tau) \nonumber \\
&= e^{-i\hat{H}_m\tau} \hat{\rho}_{NESS}^{\rm PReB}\prod_\ell\frac{e^{-\beta_\ell(\hat{H}_{B_{\ell m}}-\mu_\ell \hat{N}_{B_{\ell m}})}}{Z_{B_{\ell m}}} e^{i\hat{H}_m\tau},
\end{align}
such that ${\rm Tr}_B(\hat{\rho}_{\rm tot}^{NESS}(\tau))=\hat{\rho}_{NESS}^{\rm PReB}$.
In terms of $\hat{\rho}_{\rm tot}^{NESS}(\tau)$, the power and the heat dissipation rates into the baths are given by
\begin{align}
\label{PReB_NESS_thermodynamics}
& P_{\rm ext}=-\sum_{\ell}\dot{H}_{SB_\ell}, \nonumber \\
&P_{\rm chem}=-\sum_{\ell} \mu_\ell\dot{N}_{B_\ell}, \\
&\dot{Q}_{\ell}=\dot{H}_{B_\ell} - \mu_\ell \dot{N}_{B_\ell}, ~~\textrm{where} \nonumber \\
& \dot{H}_{SB_\ell}\nonumber \\
&= \frac{{\rm Tr}\left(\hat{H}_{SB_{\ell}} \hat{\rho}_{\rm tot}^{NESS}(\tau)\right)-{\rm Tr}\left(\hat{H}_{SB_{\ell}} \hat{\rho}_{\rm tot}^{NESS}(0)\right)}{\tau}, \nonumber \\
& \dot{N}_{B_\ell} \nonumber \\
&=\frac{{\rm Tr}\left(\hat{N}_{B_{\ell}} \hat{\rho}_{\rm tot}^{NESS}(\tau)\right)-{\rm Tr}\left(\hat{N}_{B_{\ell}} \hat{\rho}_{\rm tot}^{NESS}(0)\right)}{\tau}, \nonumber \\
& \dot{H}_{B_\ell} \nonumber \\
&= \frac{{\rm Tr}\left(\hat{H}_{B_{\ell}} \hat{\rho}_{\rm tot}^{NESS}(\tau)\right)-{\rm Tr}\left(\hat{H}_{B_{\ell}} \hat{\rho}_{\rm tot}^{NESS}(0)\right)}{\tau}. \nonumber
\end{align}
In the above, $\dot{N}_{B_\ell}$, $\dot{H}_{B_\ell}$ and $\dot{Q}_\ell$ are the cycle-averaged particle, energy and heat currents into the $\ell$th bath at the NESS of PReB. We consider cases where the Hamiltonian $\hat{H}_m$ conserves the total number of particles. Using this property, at the NESS of PReB, we have 
\begin{align}
\label{number_conservation_NESS_PReB}
\sum_\ell \dot{N}_{B_\ell}=0.
\end{align}
 This shows that, if there is no chemical potential difference between the baths, $\mu_\ell=\mu$, there is no chemical power, $P_{\rm chem}=-\mu \sum_{\ell}\dot{N}_{B_\ell}=0$, consistent with physical intuition. However, it is interesting to note that, due to the external power from the switching of the system-bath couplings, $\sum_\ell \dot{H}_{B_\ell}\neq 0$. In fact, by using the definitions of $P_{\rm chem}$ and $\dot{Q}_\ell$ in the statement of first law, Eq. \eqref{first_law_NESS}, we see that  $P_{\rm ext}=\sum_\ell \dot{H}_{B_\ell} $.

The above intuitive definitions of power and heat currents can be obtained starting from a more microscopic and general description for thermodynamics of open quantum systems~\cite{Esposito_2010,Reeb_2014,Landi_2021,Strasberg_tutorial_2021}, given in Appendix~\ref{Appendix:general_definitions_for_thermodynamics}. This microscopic general description crucially requires that the baths have infinite degrees of freedom, which, as we have discussed before, is also required for realizing the PReB process. It however does not have any restriction on strength of system-bath couplings or the speed of driving (which here corresponds to rate of switching of system-bath couplings). Moreover, due to the particular nature of the PReB process, even though the presence of macroscopic baths are important, all thermodynamic quantities can be calculated by using one copy of the finite-size tight-binding chains for each bath, as we see above at the NESS. This is shown in general, for all times, in Appendix~\ref{Appendix:PReB_thermo}, where the derivation of the expressions in Eq.~(\ref{PReB_NESS_thermodynamics}) starting from the general description for the thermodynamics of open quantum systems in Appendix~\ref{Appendix:general_definitions_for_thermodynamics} is also discussed.

\subsection{NESS of continuous-time autonomous process: $n=1$, $\tau \rightarrow \infty$}

As discussed before, the NESS of the continuous-time autonomous process is obtained by taking $n=1$ and $\tau\rightarrow \infty$. The above expressions for power and heat dissipation rate into the baths, with this substitution holds in that case. Since the system-bath coupling energy corresponds to the expectation value of a local operator of the set-up, it cannot diverge. Consequently, power extraction from switching-off the system bath coupling goes to zero as $\tau\rightarrow \infty$. The currents from the baths become constant, thereby giving, for the continuous time autonomous process,
\begin{align}
\label{continuous_time_NESS_thermodynamics}
& P_{\rm ext}^{\rm cont}=0,~~P^{\rm cont}=P_{\rm chem}^{\rm cont}=\sum_{\ell} \mu_\ell I_{B_\ell \rightarrow S},\nonumber \\
&\dot{Q}_{\ell}^{\rm cont}=- \left[J_{B_\ell \rightarrow S}-\mu_\ell  I_{B_\ell \rightarrow S}\right], \\
& P_{\rm chem}^{\rm cont}=\sum_\ell \dot{Q}_\ell^{\rm cont},~~\sigma^{\rm cont} = \sum_\ell \beta_\ell \dot{Q}_\ell^{\rm cont}\geq 0 \nonumber
\end{align}
where $J_{B_\ell \rightarrow S}$ is the energy current into the system from the $\ell$th bath at NESS, and $I_{B_\ell \rightarrow S}$ is the particle current into the system from the $\ell$th bath at the NESS.  
Similar to the NESS of PReB, the conservation of the number of particles gives $\sum_{\ell} I_{B_\ell \rightarrow S}=0$, which shows that without any chemical potential difference, $\mu_\ell=\mu$, there is no chemical power, $P_{\rm chem}^{\rm cont}=\mu\sum_{\ell} I_{B_\ell \rightarrow S}=0$. However, unlike in the NESS of PReB, since in this case there is no external power, we have $\sum_\ell J_{B_\ell \rightarrow S} =0$.

The above discussion also shows that with increasing $\tau$, the NESS thermodynamic quantities of PReB will converge to those of the continuous-time autonomous process. 

\section{Obtaining the NESS for Gaussian systems}
\label{Sec:Lyapunov}

In order to explore the thermodynamics at the NESS, we first need to find the NESS of PReB. For a general quantum many-body system, this is a numerically difficult, but not impossible, problem. The difficulty primarily arises because the Hilbert space dimension of the whole set-up scales exponentially with size of the set-up. However, the type of system we have considered in the example here falls under the special class of Gaussian systems. The Hamiltonian governing the process during the $m$th step  can be written in the form
\begin{align}
\label{full_set_up_non_interacting}
\hat{H}_m=\sum_{p,q=1}^{L_S+2L_B} \mathbf{H}_{p,q} \hat{d}_p^\dagger \hat{d}_q,
\end{align}
where $\hat{d}_p$ is the fermionic annihilation operator of either a system or an environment site, and $L_S$ is the number of sites in the system ($L_S=2$ in our example), and $L_B$ length of the finite-size chain representing the bath obtained by chain-mapping, see Eqs.~(\ref{chain_mapping}) and (\ref{chain_size}). Here, $\mathbf{H}$ is a $(L_S~+~2L_B) \times (L_S+2L_B)$ dimensional real symmetric matrix, often called the single-particle Hamiltonian. 
The matrix $\mathbf{H}$ can be written in the following block form,
\begin{align}
\label{single_particle_H}
\mathbf{H}=\left[
\begin{array}{ccc}
\mathbf{H}_S & \mathbf{H}_{SB_1} & \mathbf{H}_{SB_2} \\
\mathbf{H}_{SB_1}^T & \mathbf{H}_{B_1} & \mathbf{0} \\
\mathbf{H}_{SB_2}^T & \mathbf{0} & \mathbf{H}_{B_2}
\end{array}
\right].
\end{align}
Here, the $L_S \times L_S$ matrix $\mathbf{H}_{S}$ is the single-particle Hamiltonian of the system, the $L_S \times L_B$ matrix $\mathbf{H}_{SB_1}$ ($\mathbf{H}_{SB_2}$) gives the coupling between the system and the first (second) bath, $\mathbf{H}_{SB_1}^T$ ($\mathbf{H}_{SB_2}^T$) is its transpose, $\mathbf{H}_{B_1}$ ($\mathbf{H}_{B_2}$) gives the single particle Hamiltonian of the first (second) finite-size environment.
As shown in Appendix~\ref{Appendix:Gaussian_systems}, for such systems, it is possible to cast the problem in terms of matrices that scale linearly with the size of the set-up. The PReB process can be described in terms of the correlation matrix of the system, sometimes called the single-particle density matrix,
\begin{align}
\mathbf{C}_{S_{p,q}}(t)={\rm Tr}\left(\hat{\rho}_{\rm tot}(t) \hat{c}_p^\dagger \hat{c}_q \right).
\end{align}
The correlation matrix of the system after the $m+1$th step of the PReB process is given by (see Appendix~\ref{Appendix:PReB_from_correlation_matrix})
\begin{align}
\label{Gaussian_PReB_system}
& \mathbf{C}^{(m+1)}_S = \mathbf{G}_S^\dagger(\tau) \mathbf{C}^{(m)}_S \mathbf{G}_S(\tau) + \mathbf{P}_S(\tau), \nonumber \\
& \mathbf{P}_S(\tau)=\mathbf{G}_{B_1 S}^\dagger(\tau) \mathbf{C}_{B_1}^{\rm therm}\mathbf{G}_{B_1 S}(\tau) \nonumber \\
&\qquad \hspace*{10pt} + \mathbf{G}_{B_2 S}^\dagger(\tau) \mathbf{C}_{B_2}^{\rm therm}\mathbf{G}_{B_2 S}(\tau).
\end{align} 
where $\mathbf{G}_S$, $\mathbf{G}_{B_1 S}$, $\mathbf{G}_{B_2 S}$ are defined via the identification,
\begin{align}
e^{-i\mathbf{H}\tau} = \left[
\begin{array}{ccc}
\mathbf{G}_S(\tau) & \mathbf{G}_{SB_1}(\tau) & \mathbf{G}_{SB_2}(\tau) \\
\mathbf{G}_{B_1 S}(\tau) & \mathbf{G}_{B_1}(\tau) & \mathbf{G}_{B_1 B_2}(\tau) \\
\mathbf{G}_{B_2 S}(\tau) & \mathbf{G}_{B_1 B_2}(\tau) &  \mathbf{G}_{B_2}(\tau)
\end{array}
\right],
\end{align}
with $\mathbf{H}$ as given in Eq.(\ref{single_particle_H}).
The $L_S \times L_S$ matrix $\mathbf{G}_S(\tau)$ in the above equation is nothing but the time-domain retarded non-equilibrium Green's function (NEGF) of the system in the presence of the finite-size environments, evaluated at time $\tau$ \cite{Jauho_book,Dhar_Sen_2006}.
The discrete-time dynamics governed by an equation of the above form can have a unique NESS if and only if all eigenvalues of $\mathbf{G}_S(\tau)$ have magnitude less than $1$. At the NESS, each step of the PReB process leaves the correlation matrix of the system invariant. Thus, from Eq.(\ref{Gaussian_PReB_system}), it must be the solution of the following matrix equation,
\begin{align}
\label{discrete_Lyapunov}
\mathbf{C}^{NESS}_S -\mathbf{G}_S^\dagger(\tau) \mathbf{C}^{NESS}_S \mathbf{G}_S(\tau)=\mathbf{P}_S(\tau),
\end{align} 
where $\mathbf{C}^{NESS}_S$ is the correlation matrix of the system at the NESS. This equation has the form of the discrete-time Lyapunov equation, which is well-studied in mathematics and engineering~\cite{Control_theory_book1,Control_theory_book2}. Given the matrices $\mathbf{G}_S(\tau)$ and $\mathbf{P}_S(\tau)$ there exists efficient ways to solve the discrete-time Lyapunov equation, using existing routines in standard languages for scientific programming (e.g. python, matlab and mathematica). This therefore allows us to directly find the NESS of the PReB process for Gaussian systems without explicitly carrying out the time-evolution.

Once we have the NESS correlation matrix of the system, we can calculate the thermodynamic quantities at the NESS defined in Eq.~(\ref{PReB_NESS_thermodynamics}) as follows. We first define the correlation matrix of the set-up at the NESS,
\begin{align}
\mathbf{C}^{NESS} = \left[
\begin{array}{ccc}
\mathbf{C}_S^{NESS} & \mathbf{0} & \mathbf{0} \\
\mathbf{0} & \mathbf{C}_{B_1}^{\rm therm} & \mathbf{0} \\
\mathbf{0} & \mathbf{0} & \mathbf{C}_{B_2}^{\rm therm}
\end{array}
\right],
\end{align}
where $\mathbf{C}_{B_\ell}$, $\ell=\{1,2\}$ are the thermal correlation matrices of the finite-size chains (see Eq.~(\ref{chain_mapping})) that are connected to the system during one step of the PReB process. Then we evolve this matrix for a time $\tau$,
\begin{align}
\mathbf{C}^{NESS}(\tau) = e^{i\mathbf{H} \tau} \mathbf{C}^{NESS} e^{-i\mathbf{H} \tau}.
\end{align}
The various thermodynamic quantities in Eq.~(\ref{PReB_NESS_thermodynamics}) correspond to the sum of various elements of $\mathbf{C}^{NESS}(\tau)$. Thus, we see that we can also obtain all the required thermodynamic quantities at the NESS directly using finite-size chains without explicitly carrying out the full time evolution. Standard NEGF results are available for the NESS of the continuous-time dynamics~\cite{Jauho_book,Dhar_Sen_2006,Dhar_2012}, which can be used for comparison in the large $\tau$ limit. This makes exploration of thermodynamics for the NESS much simpler.

We remark that the entire formalism in this section can be straightforwardly generalized to other Gaussian systems, such as those composed of harmonic oscillators. Furthermore, the dynamics of any collisional or repeated interaction model  involving only Gaussian states (including those involving athermal Gaussian environment states, for example, \cite{Grimmer_2018,Camasca_2021,Leitch_2021}) can be cast in an analogous language. In addition, these results remain true for systems with an arbitrary number of sites, in a lattice of arbitrary geometry and dimension. Here, however, we will use it for our simple two-site example.

\begin{figure*}
\includegraphics[width=\textwidth]{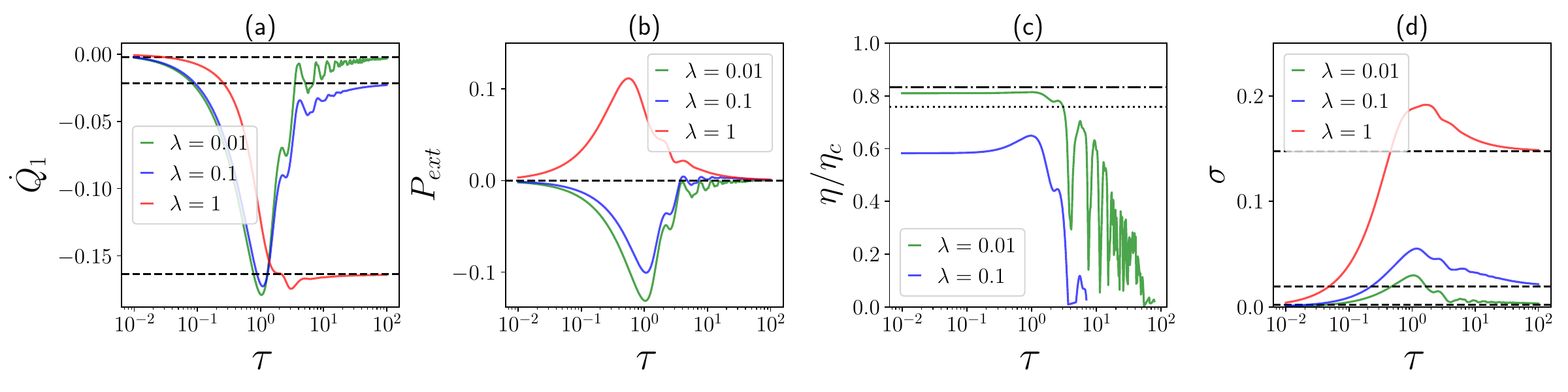}
\caption{\label{fig:heat_engine} {\bf Heat engine at NESS of PReB:} {\bf (a)} Heat dissipation rate into the hot bath $\dot{Q}_1$ (negative value means heat flows into the system). {\bf (b)} the external power input from switching on and off system-bath couplings $P_{ext}$ (negative value means work is done by the system), {\bf (c)} the efficiency in the heat engine regime, i.e, when $P_{ext}<0$, {\bf (d)} the entropy production rate at NESS of PReB process, as a function of $\tau$, for various values of $\lambda$. Note that the chemical power $P_{chem}=0$ since there is no chemical potential bias. The horizontal dashed lines in (a), (b) and (d) give results for NESS in the $\tau\rightarrow \infty$ limit which corresponds to the continous-time autonomous case. The horizontal dash-dotted line in (c) marks $\eta/\eta_c$ in the $\lambda\rightarrow 0$ limit which is independent of $\tau$. The horizontal dotted line in (c) shows the value for Curzon-Ahlborn efficiency at the same temperatures. Parameters: $\beta_1=0.1$, $\beta_2=1$, $\mu=-2$, and other parameters as given in Eq.(\ref{fixed_parameters}). All energy parameters are in units of the system hopping parameter $g$, while all time parameters are in units of $g^{-1}$.}
\end{figure*}

\section{The heat engine and refrigerator regimes}
\label{Sec:heat engine and refrigerator regimes}

In the heat engine regime, power is extracted from the system by using the heat taken from the hot bath and therefore,
\begin{align}
\label{heat_engine}
P=P_{\rm ext}+P_{\rm chem}<0,~~\dot{Q}_{1} <0,
\end{align} 
(see Table~\ref{sign_convention} for sign convention.) The efficiency $\eta$ of the heat engine is then the ratio of extracted power over the rate of heat flow from the hot bath, which is upper bounded by the Carnot efficiency $\eta_c$,
\begin{align}
\eta = \frac{P}{\dot{Q}_1},~~\eta_c = 1-\frac{\beta_1}{\beta_2},~~\eta \leq \eta_c
\end{align}
In the refrigerator regime, heat is extracted from the cold bath, thereby cooling it further utilizing the power input into the system and therefore,
\begin{align}
\label{refrigerator}
\dot{Q}_{2} <0,~~P=P_{\rm ext}+P_{\rm chem} > 0.
\end{align}
The coefficient of performance ($COP$) of the refrigerator is then the ratio of cooling rate over input power, which is upper bounded by that of the Carnot refrigerator,
\begin{align}
&COP = \frac{-\dot{Q}_{2}}{P},~~COP \leq COP_c,\nonumber \\
& COP_c = \frac{1}{\beta_2/\beta_1-1},
\end{align}
where $COP_c$ is the coefficient of performance of the Carnot refrigerator.
The bound on $\eta$ and $COP$ follow directly from the expressions for the first and the second law of thermodynamics, Eqs. (\ref{first_law_NESS}), (\ref{second_law_NESS}), as discussed in Appendix~\ref{Appendix:heat_engines_and_refrigerators}.

In our numerical simulations, we set $g=1$ as the unit for energy. For simplicity, we fix the following parameters
\begin{align}
\label{fixed_parameters}
\kappa=2,~~\omega_{01}=2,~~\omega_{02}=-1,~~\Lambda=6,
\end{align}
and explore the thermodynamics at the NESS of PReB by varying the parameters, $\beta_1, \beta_2,\mu, \tau, \lambda$ (see Eqs.\eqref{Lorentzian}, \eqref{temperature_and_chemical_potential}). For each value of $\tau$, we map the bath spectral functions given in Eq.\eqref{Lorentzian} to chains of size $L_B$, see Eq.\eqref{chain_size}, and use the formalism of Sec.~\ref{Sec:Lyapunov} to obtain the results numerically. Recall that since there is no chemical potential bias, there is no chemical power. The only source of power is the switching of system-bath couplings. Consequently, there can be no QTM in the continuous-time autonomous limit of $\tau \to \infty$. 

\begin{figure*}
\includegraphics[width=\textwidth]{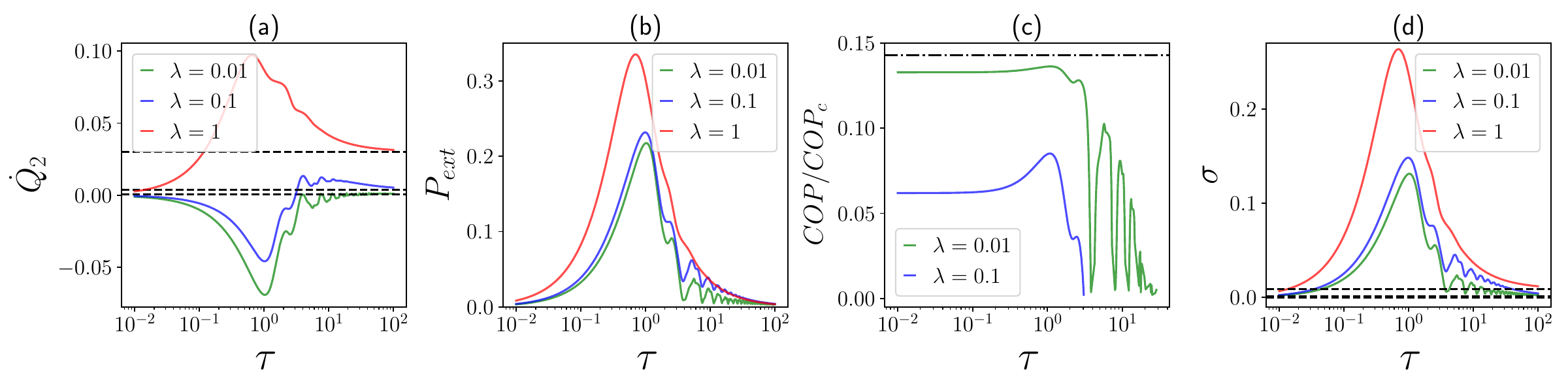}
\caption{\label{fig:refrigerator} {\bf Refrigerator at NESS of PReB:} {\bf (a)} Heat dissipation rate into the cold bath $\dot{Q}_2$ (negative value means heat flows into the system), {\bf (b)} the external power input from switching on and off system-bath couplings $P_{ext}$, {\bf (c)} the coefficient of performance in the refrigerating regime, i.e, when $\dot{Q}_2<0$.  {\bf (d)} the entropy production rate at NESS of PReB process, as a function of $\tau$, for various values of $\lambda$. Note that the chemical power $P_{chem}=0$ since there is no chemical potential bias. The horizontal dashed lines in (a), (b) and (d) give results in the $\tau \rightarrow \infty$ limit, which corresponds to continous-time autonomous case. The horizontal dash-dotted line in (c) marks $COP/COP_c$ in the limit of $\lambda\rightarrow 0$ which is independent of $\tau$. Parameters: $\beta_1=0.7$, $\beta_2=1$, $\mu=-2$, and other parameters as given in Eq.(\ref{fixed_parameters}). All energy parameters are in units of the system hopping parameter $g$, while all time are parameters in units of $g^{-1}$.}
\end{figure*}

Representative plots of thermodynamic quantities as a function of $\tau$ for a fixed choice of temperatures and chemical potentials and for three different values of $\lambda$ are shown in Fig.~\ref{fig:heat_engine}.  We see in Figs.~\ref{fig:heat_engine}(a),(b) and (d) that the thermodynamic quantities converge to those of the NESS of the continuous-time autonomous process for increasing $\tau$, as expected. For $\tau \ll g^{-1}$, all currents, and hence the entropy production rate,  decrease upon reducing $\tau$. This is the Zeno-like regime where transport is slowed down because the baths are refreshed much faster than the system time scale, $g^{-1}$.  We see in Fig.~\ref{fig:heat_engine}(a) that $\dot{Q}_1$ is always negative which means heat flows out of the hot bath. At small values of $\lambda$, the heat flow, and correspondingly the entropy production rate, is small in the large $\tau$ limit. This is because the connection with the residual bath decreases with $\lambda$, as can be seen from Eq.~(\ref{rc_mapping_Lorentzian}). For intermediate values of $\tau$,  the heat flow from the hot bath can be much larger. In Fig.~\ref{fig:heat_engine}(b), we see that at such intermediate values of $\tau$, we can have $P_{ext}<0$. Hence, the power associated with switching on and off the system-bath couplings can be extracted, thereby realizing a unique type of heat engine. This effect happens for $\lambda \ll g$ but is lost as $\lambda$ is increased, clearly reflecting the role of the structure of the bath-spectral functions.  The efficiency of the heat engine regime is shown as a function of $\tau$ in Fig.~\ref{fig:heat_engine}(c). On decreasing $\tau$ from a large value, the efficiency first shows fluctuations, and then reaches a plateau for small values of $\tau$. The horizontal dash-dotted line in Fig.~\ref{fig:heat_engine}(c) shows the efficiency in the limit of $\lambda \rightarrow 0$, which we will discuss in detail in the next subsection. The horizontal dotted line in Fig.~\ref{fig:heat_engine}(c) shows the Curzon-Ahlborn prediction for efficiency at maximum power, $\eta_{CA}=1-\sqrt{\frac{\beta_1}{\beta_2}}$. Although originally derived for the endoreversible Carnot cycle~\cite{CA}, this value is sometimes thought to upper bound efficiency at maximum power more generally. We see that here this is clearly not the case. This is not surprising because the Curzon-Ahlborn prediction is non-universal and has been shown be violated in other quantum heat engines also (see, for example, Ref.\cite{Myers_2022} for more details).

In Fig.~\ref{fig:refrigerator}, we plot the thermodynamic quantities as a function of $\tau$ for a different choice of temperatures and chemical potential of the baths, for three values of $\lambda$. We again see that for increasing $\tau$, all thermodynamic quantities converge to those of the continuous-time autonomous NESS, while for $\tau \ll g$, all currents decrease with decreasing $\tau$. However, here we find that $P_{ext}>0$ for all values of $\tau$. More interestingly, $\dot{Q}_{2}$, which is shown in Fig.~\ref{fig:refrigerator}(a), becomes negative at intermediate values of $\tau$ for $\lambda \ll g$, thereby realizing a unique type of refrigerator. This refrigerator utilizes the power from the switching of system-bath couplings to cool the cold bath. The dependence on $\lambda$ once again highlights the effect of the details of bath spectral functions. In Fig.~\ref{fig:refrigerator}(c), we show the $COP$ in the refrigerating regime. The $COP$ of the refrigerator shows a similar behavior to the efficiency of the heat engine as a function of $\tau$. On reducing $\tau$ from $\tau>g^{-1}$, first it shows fluctuations above an overall increase, and then for $\tau \ll g^{-1}$ the $COP$ saturates. The horizontal dash-dotted line in  Fig.~\ref{fig:refrigerator}(c) corresponds to the $COP$ in the limit $\lambda \rightarrow 0$ which we will discuss in detail in the next subsection.

An extremely interesting point to note from Fig.~\ref{fig:heat_engine} (Fig.~\ref{fig:refrigerator}), is that $\eta$ ($COP$), extracted power $-P_{ext}$ (cooling rate $-\dot{Q}_2$) and entropy production rate $\sigma$ are all maximized at $\tau\sim O(g^{-1})$. Thus, as a function of $\tau$, these QTMs become most efficient when the system is most irreversible. This particular property is in stark contrast with traditional cyclic heat engines and refrigerators which become most efficient in the infinitely slow regime where the system becomes most reversible~\cite{Leff_1975,Leff_2018,Myers_2022}. This regime would have power going to zero, therefore leading to a trade-off between extracted power and efficiency of a heat engine as a function of cycle duration, $\tau$. However, since the heat engine and the refrigerator in our case become most efficient at the value of $\tau$ where they are most irreversible, there is no trade-off between efficiency and power as function of $\tau$.

We see in Fig.~\ref{fig:heat_engine}(c) (Fig.~\ref{fig:refrigerator}(c)), for $\tau \lesssim g^{-1}$, as $\lambda$ is reduced, the efficiency ($COP$) increases towards its value in the $\lambda \rightarrow 0$ limit. The maximum extracted power (maximum cooling rate) also increases on reducing $\lambda$, showing no trade-off as a function of $\lambda$ also.  
This prompts a deeper look into the $\lambda \rightarrow 0$ limit, which also allows us to find necessary conditions for realization of the heat engine and refrigerating regimes, as we show below.

\section{Necessary conditions for  realizing the heat engine and the refrigerator regimes}
\label{Sec: conditions for heat engine and refrigerator regimes}

In the preceding section, we have seen that the heat engine and the refrigerating regimes in the absence of any chemical potential bias are realized for intermediate values of $\tau$ when $\lambda$ is small. For $\lambda\rightarrow 0$, as explained before in Eqs.~(\ref{rc_mapping_Lorentzian}), (\ref{tau_R_Lorentizian}), the set-up is reduced to a standard collisional model given by repeated interactions with infinite copies of single-site fermionic environments. In this case, it can be seen that the NESS cycle-averaged heat currents into the baths (see Eq.~\eqref{PReB_NESS_thermodynamics}) become proportional to the corresponding particle currents,
\begin{align}
\dot{Q}_{\ell}=(\omega_{0\ell}-\mu)\dot{N}_{B_\ell}~~({\rm for}~~\lambda \rightarrow 0).
\end{align}
This is the so-called tight-coupling condition. Under this condition, using Eq.~(\ref{number_conservation_NESS_PReB}) we see that,
\begin{align}
\frac{\dot{Q}_{1}}{\dot{Q}_{2}}=-\frac{\omega_{01}-\mu}{\omega_{02}-\mu}~~{\rm for}~~\lambda \rightarrow 0.
\end{align} 
Using this result with the first law of thermodynamics Eq.~(\ref{first_law_NESS}) gives the following expression for the efficiency
\begin{align}
\label{collisonal_efficiency}
\eta=1-\frac{\omega_{02}-\mu}{\omega_{01}-\mu}~~{\rm for}~~\lambda \rightarrow 0,
\end{align}
irrespective of any further details of the system. Similarly the $COP$ for the refrigerating regime is given by
\begin{align}
\label{collisonal_COP}
COP = \frac{1}{(\omega_{01}-\mu)/(\omega_{02}-\mu)-1}~~{\rm for}~~\lambda \rightarrow 0,
\end{align}
irrespective of any further details about the system. The dash-dotted lines in Figs.~\ref{fig:heat_engine}(c) and ~\ref{fig:refrigerator}(c) are calculated using the above results for $\eta$ and $COP$. Clearly, we see that this value is approached from below for $\tau \lesssim g^{-1}$ with decreasing $\lambda$. Noting that the efficiency and the COP are positive and upper bounded by those of the Carnot cycle,  we arrive at the following necessary (but not sufficient) conditions 
\begin{align}
\label{heat_engine_condition}
&\frac{\beta_1}{\beta_2}< \frac{\omega_{02}-\mu}{\omega_{01}-\mu}<1~~\textrm{for the heat engine regime,}  \\
\label{refrigeration_condition}
& 0< \frac{\omega_{02}-\mu}{\omega_{01}-\mu}< \frac{\beta_1}{\beta_2}~~\textrm{for the refrigerating regime.}
\end{align}
Although calculated in the $\lambda \to 0$ limit, we expect these results to hold even when $\lambda$ is small (but non-zero) and $\tau\ll \lambda^{-1}$. This is because under such conditions, to a very good approximation, at best the first site of the bath obtained after chain-mapping can be affected, while the remaining sites are almost unaffected. Thus, for $\tau\sim O(g^{-1})$ and $\lambda \ll g$, where the our QTM gives best performance, the single-site approximation should hold. However,  if $\tau> \lambda^{-1}$ the single-site approximation will not hold even for a small $\lambda$. Rather, on increasing $\tau$, the behavior will converge towards the continuous-time autonomous result, where there can be no QTM without a chemical potential bias. Interestingly, Eqs.~(\ref{heat_engine_condition}) and (\ref{refrigeration_condition}) are similar to conditions obtained in the context of the thermodynamics of local Lindblad equations in Ref.~\cite{Gabriele_NJP_2018}.

\begin{figure}
\centering
\includegraphics[width=\columnwidth]{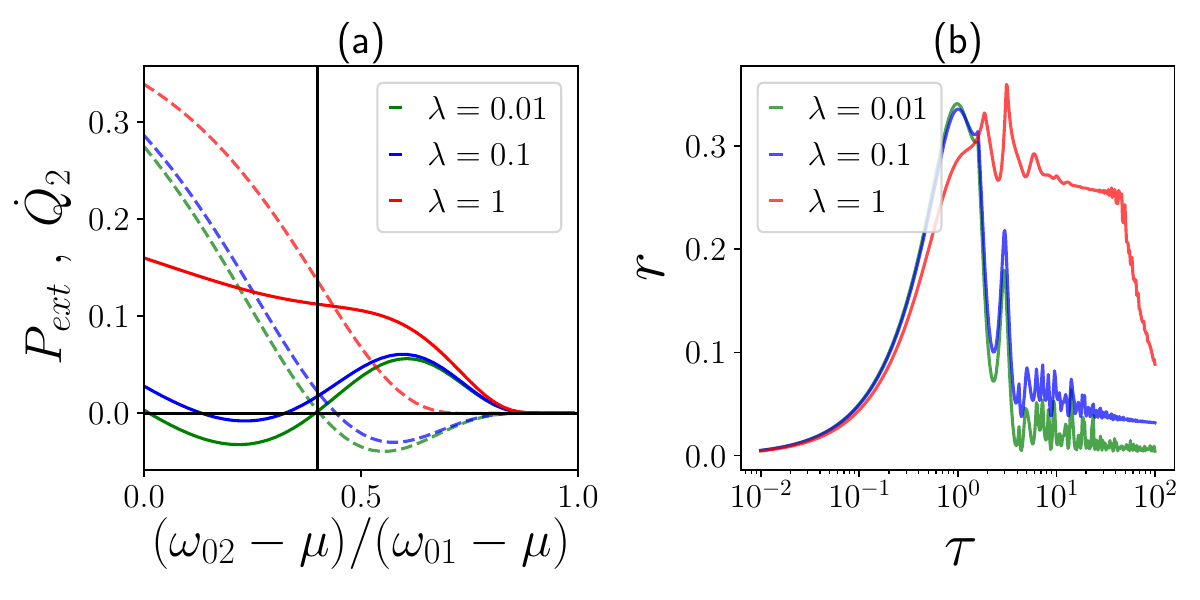}
\caption{\label{fig:machine_condition_and_t_NESS} {\bf (a) Necessary condition for heat engine and refrigerating regimes:} The figure shows plot of power from external driving $P_{ext}$ (dashed lines, negative values correspond to heat engine regime) and heat dissipated into cold bath $\dot{Q}_2$ (continuous lines, negative values correspond to refrigerator regime) as a function of $(\omega_{02}-\mu)/(\omega_{01}-\mu)$ at the NESS of the PReB process with $\tau=1$. The black continuous vertical line corresponds to $\beta_1/\beta_2$, with $\beta_1=0.4$, $\beta_2=1$. Here $\omega_{01}$, $\omega_{02}$ are fixed at chosen values (see Eq. (\ref{fixed_parameters})) and only $\mu$ has been varied to obtain the plot. {\bf (b) Zeno and anti-Zeno-like regimes:} The figure shows the rate of approach to NESS for the PReB process, $r$, as a function of $\tau$. All energy parameters are in units of the system hopping parameter $g$, while all time parameters are in units of $g^{-1}$.    }
\end{figure}

We numerically check the above discussion in our set-up. In Fig.~\ref{fig:machine_condition_and_t_NESS}(a), we plot $P_{ext}$ and $\dot{Q}_2$ for $\tau=g^{-1}$ as a function of the ratio in Eqs.~(\ref{heat_engine_condition}) and (\ref{refrigeration_condition}), for various values of $\lambda$. We clearly see that negative values of $P_{ext}$ exist only when Eq.~(\ref{heat_engine_condition}) is satisfied, and negative values of $\dot{Q}_2$ exist only when Eq.~(\ref{refrigeration_condition}) is satisfied. Also, on increasing $\lambda$ the possibility of having negative values for either quantities disappear. In obtaining Fig.~\ref{fig:machine_condition_and_t_NESS}(a), only $\mu$ was varied while $\omega_{01}$ and $\omega_{02}$ were fixed at the chosen values given in Eq.~(\ref{fixed_parameters}). This shows that even though the chemical potentials are the same in our example, they play a non-trivial role in the realization of the heat engine and refrigerating regimes.

Although there is no trade-off between power and efficiency of the heat engine as a function of $\tau$ and $\lambda$, there is actually a trade-off if we want to push the efficiency in the limit $\lambda \rightarrow 0$ to the Carnot efficiency. From Eq.~(\ref{collisonal_efficiency}), we see that this is reached only when $(\omega_{02}-\mu)/(\omega_{01}-\mu)= \beta_1/\beta_2$, which corresponds to the boundary of heat engine and refrigerating conditions in Eqs.~(\ref{heat_engine_condition}), (\ref{refrigeration_condition}). In this case, as is evident in Fig.~\ref{fig:machine_condition_and_t_NESS}(a), both $P_{ext}$ and $\dot{Q}_2$ go to zero in the limit $\lambda \rightarrow 0$. The entropy production rate can also be shown to go to zero in this case. In this sense, we recover the usual trade-off seen on approaching the Carnot limit. Nevertheless, as we have seen in Fig.~\ref{fig:heat_engine}(c), we can reach quite close to the Carnot limit at finite power. Also note that, from Fig.~\ref{fig:machine_condition_and_t_NESS}(a) the efficiency at maximum output power in the $\lambda \rightarrow 0$ limit can be estimated to be $0.757 \eta_c$, which is significantly greater than the Curzon-Ahlborn value $0.612 \eta_c$ at the chosen temperatures.

\section{Zeno-like and anti-Zeno-like effects}
\label{Sec: Zeno and anti-Zeno}

We have previously remarked on the Zeno-effect-like behavior of the set-up when $\tau \ll g^{-1}$. We have also seen that the magnitude of heat currents can increase with a decrease in $\tau$ for $\tau \gtrsim O(g^{-1})$. This increase in heat transfer is akin to what has been previously discussed as the anti-Zeno effect~\cite{Kofman_2000,Kofman_2001,Kofman_2004,Erez_2008,Alvarez_2010,Rao_2011,
Mukherjee_2020,Das_2020}, which says that repeated monitoring below a certain rate can speed up the system dynamics. In this subsection we directly check this effect. To this end, we look at the rate of approach to the NESS of the PReB process. This rate can be elegantly estimated from the eigenvalue of largest magnitude of $\mathbf{G}_S(\tau)$ in Eq.~(\ref{Gaussian_PReB_system}),
\begin{align}
r = -\frac{\log( |\mathfrak{g}_{max}|)}{\tau},
\end{align}
where $\mathfrak{g}_{max}$ is the largest magnitude eigenvalue of $\mathbf{G}_S(\tau)$. The NESS of PReB is effectively reached when $n\tau \gg r^{-1}$. Since $\mathbf{G}_S(\tau)$ is independent of the temperatures and chemical potentials of the baths, this rate of approach to the NESS is also independent of those parameters. In Fig.~\ref{fig:machine_condition_and_t_NESS}(b) we plot the rate $r$ as a function of $\tau$ for various values of $\lambda$. For $\tau < g^{-1}$, for decreasing $\tau$ we see a monotonic decrease in $r$. Thus, the dynamics slows down with an increase in the rate of refreshing of the baths. This is the Zeno-like regime. Comparing with Fig.~\ref{fig:heat_engine} (Fig.~\ref{fig:refrigerator}), we see that in the Zeno-like regime, the efficiency (coefficient of performance) saturates, while the power (cooling rate) goes to zero with  a decrease in $\tau$. For $\tau> g^{-1}$, we find a non-monotonic dependence of $r$ on $\tau$. In fact for $\tau \gg g^{-1}$, we find that, although there is non-monotonicity, there is an overall decay. Thus, in this regime, increasing the rate of refreshing of the baths can speed up the dynamics, as evidenced by an increase in rate of approach to the NESS with decrease in $\tau$. This is exactly akin to the anti-Zeno effect~\cite{Kofman_2000,Kofman_2001,Kofman_2004,Erez_2008,Alvarez_2010,Rao_2011,
Mukherjee_2020,Das_2020}. It is in this regime that power and heat currents (and hence the entropy production rate) show a simultaneous increase in magnitude for a decrease in $\tau$. The crossover into the Zeno-like regime occurs at $\tau \sim g^{-1}$ and it is at the same order of $\tau$ that the magnitude of all these quantities are maximized. Thus, their simultaneous increase can be linked with a speed-up of the dynamics due to the anti-Zeno-like effect. Anti-zeno advantage in QTMs have been reported previously in completely different settings \cite{Mukherjee_2020,Das_2020}.  While the rich variation of $r$ with $\tau$ is interesting, we defer a detailed study of this behavior to future work. Instead, next we look at the cost of increasing efficiency and power simultaneously.

\section{Trade-off in terms of complexity of realization}
\label{Sec:Trade-off}

We have seen that the efficiency and power of the heat engine regime can be maximized together as function of $\tau$ and $\lambda$ without any trade-off. However, Eq.~(\ref{tau_R_Lorentizian}) shows that there is a different kind of cost for increasing efficiency and power (cooling rate) of these thermal machines. Namely, the number of copies $\mathfrak{N}$ of the baths required to realize the PReB process with such parameters increases. The number of copies $\mathfrak{N}$ is minimum in the continuous-time autonomous limit, $\tau\rightarrow \infty$. However, in the absence of a chemical potential difference, there can be no quantum thermal machine in such a limit. For $\tau\sim g^{-1}$, where we have found the most efficient QTMs, a much larger value of $\mathfrak{N}$ is needed. Moreover, we also need a small value of $\lambda$, which further increases the value of $\mathfrak{N}$. Thus, the increase in performance of the heat engine and refrigerator regimes with PReB come at the cost increasing the complexity of realizing such devices at the corresponding parameters.

\section{Where is the heat extracted form?}
\label{Sec: where is the heat extracted form}

Any QTM requires extraction of heat from a thermal bath. However, in the PReB process the baths are periodically refreshed to their original initial states. It is therefore interesting to consider where the heat required is extracted from. For this, we need to go back to the physical realization of the PReB process, as discussed in Sec.~\ref{Sec: Physically_engineering_PReB} and as shown in Fig.~\ref{fig:schematic}. As mentioned, only a finite part of the macroscopic baths affects the dynamics of the system, while the remaining part plays the role of rethermalizing the finite part when disconnected from the system. The key point to note is that in the physical realization of the PReB process, the state of the whole macroscopic bath is not refreshed, but rather only the finite part of the bath that affects the system dynamics is effectively rethermalized.  The heat is extracted from the residual infinite degrees of freedom during the process of rethermalization. 

This once again highlights the crucial need for the baths to have infinite degrees of freedom for the physical realization of the PReB process. The infinite degrees of freedom provide an infinite capacity for heat and particles. At any finite time, only a finite part of the infinite degrees of freedom is affected. This is why the macroscopic properties, i.e. the temperatures and chemical potentials of the baths, can be considered constant at the NESS of PReB, as long as $n\tau \gg r^{-1}$ but finite. This is despite the fact that, at any finite time, microscopically, a finite part each bath is always changing.  

However, any real bath, no matter how macroscopic, would still have a finite, but extremely large number of degrees of freedom. So eventually, after an extremely long time, the approximation of bath temperatures and chemical potentials remaining constant will fail. The general theory for describing thermodynamics used here will then need to be modified to incorporate the possible changes in temperatures and chemical potentials due to finite bath-size effects \cite{Reeb_2014,Strasberg_2021,Strasberg_tutorial_2021}. Such directions will be explored in future works.

\section{Summary and outlook}
\label{Sec:conclusions}

{\it Interpolating between collisional QTMs and autonomous QTMs ---}
In this paper, we have considered the thermodynamics of the PReB process where a system is coupled to multiple thermal baths which, from the perspective of the system, are refreshed to their original thermal states at intervals of time $\tau$ \cite{PReB}. To do so, we have discussed how such a process may be realized physically. We show that the PReB process can be realized with a finite number of copies of each bath, relying on their self-rethermalization when disconnected from the system. For increasing $\tau$, the PReB process converges, both dynamically and thermodynamically, to the continuous-time autonomous process, where the baths are kept connected without any refreshing \cite{PReB}.  We have found that, if the bath spectral functions approach a Dirac-delta function, the PReB process can reduce to a standard collisional model based on repeated interactions with single-site environments \cite{Campbell_2021,Ciccarello_2021}.  At the NESS of the PReB process, we discovered that novel types of cyclic QTMs can be realized, $\tau$ being the cycle duration, which can therefore interpolate between standard collisional QTMs with single-site environments and autonomous QTMs with macroscopic baths \cite{Benenti_2017,Mark_2019}. These QTMs can utilize the work associated with switching on and off of the system-bath couplings. Depending on various parameters, including the nature of the spectral functions of the baths,  this work can both be extracted and be used to induce  cooling of the cold bath. We remark that in the standard collisional or repeated interaction framework it has been a challenge to include information about bath spectral functions. It is also important to mention that our treatment has neither any restriction on strength of system-bath coupling, nor any restriction on the rate of switching on and off of the system-bath couplings. 
\\

\noindent
{\it Harnessing irreversibility to boost performance ---}
We have demonstrated both the heat engine (Fig.~\ref{fig:heat_engine}) and the refrigerator (Fig.~\ref{fig:refrigerator}) regimes of operation of the NESS of PReB process using a simple example. This consists of two coupled fermionic sites, each coupled to its own fermionic bath, which, from the perspective of the system, are periodically refreshed to their original initial states. The bath spectral functions are chosen to be Lorentzian, which allows us to interpolate between structured spectral function with a pronounced peak and a reasonably featureless spectral function. In the limit of Lorentizian width going to zero, we recover a standard collisional model based on repeated interaction with single-site environments. 

We have specifically chosen the baths to have equal chemical potentials so that there can be no chemical work. This ensures that the only source of power is the switching on and off of the system-bath couplings. As a result, in the limit of the continuous-time autonomous process, $\tau \rightarrow \infty$, there can neither be work extraction nor cooling.  

Nevertheless, for finite $\tau$, both work extraction and cooling is possible depending on the width of the Lorentzian. A smaller width, which implies a more pronounced peak, aids the realization of the heat engine and refrigerating regimes. By considering the limit of the standard collisional model, we are able to provide necessary conditions on bath properties to realize the heat engine and the refrigerating regimes. Remarkably, as a function of $\tau$, the heat engine and the refrigerating regimes become most efficient when the process in most irreversible, as evidenced by a peak in entropy production rate. This is in stark contrast with QTMs based on traditional thermodynamic cycles which become most efficient in the infinitely slow limit, where the entropy production rate goes to zero. As a consequence, there is no trade-off between efficiency  and power  in the heat engine regime  and both can be maximized together as a function of the cycle duration $\tau$. 

This simultaneous increase of power and efficiency of the heat engine  occurs in the regime of $\tau$ where an anti-Zeno-like effect is seen. Interestingly, a different type of trade-off occurs. The complexity of realizing the PReB process, as evidenced by the number of copies of the baths required, increases for parameters where efficiency and power simultaneously increases. Nevertheless, the efficiency of the heat engine can be shown to remain upper bounded by the Carnot efficiency, and when this global maximum is approached, the power goes to zero. Analogous statements to the above also hold for the coefficient of performance and the cooling rate in the refrigerator regime.
\\

\noindent
{\it Insights for experimental realization ---}
To experimentally realize a PReB process, one needs to have some estimate of the self-rethermalization time or relaxation time $\tau_R$ of the baths and some choice of $\tau$ to know how many copies of the baths are required. Then, to experimentally obtain the thermodynamic quantities, one needs to know the temperature and the chemical potentials of the baths, and has to measure the energy and the particle currents from the baths. No other knowledge of the details of the baths is required, making the experimental realization plausible.

If some specific bath spectral functions are desired, they can also be modelled by engineering finite parts of the baths. The parameters for such design can be obtained via the chain-mapping procedure, Eqs.(\ref{chain_mapping}) and (\ref{rc_map})). For instance, in our simple example the Lorentzian spectral functions can be designed by engineering a single dissipative site as the bath, the source of dissipation playing the role of the infinite degrees of freedom. 
\\

\noindent
{\it Calculating thermodynamic quantities using finite-size environments ---}
We have seen that the baths need to have infinite degrees of freedom both for the physical realization of dynamics governed by the PReB process and for a consistent thermodynamic description. However, interestingly, for the calculation of both dynamics and thermodynamics of the PReB process, only a finite number of degrees of freedom of the baths play a role. Both dynamics and thermodynamics of the PReB process can be obtained by considering a single copy of finite-size environments, the size being approximately proportional to $\tau$. This is even more remarkable because with increase in $\tau$, both the dynamics and the thermodynamics of the PReB process converge to the continuous-time autonomous process. So, both dynamics and thermodynamics of the continuous-time autonomous process can be quite accurately obtained by recursively using finite, but large enough, environments. 
\\

\noindent
{\it The discrete-time Lyapunov equation for Gaussian processes ---}
Finally, we have found a simple way to describe the PReB process (and hence, any collisional or repeated interaction process) for any Gaussian system in a lattice of arbitrary dimension and geometry. This description is in terms of the correlation matrix, also called single-particle density matrix, whose equation of motion we find to be related to a discrete-time Lyapunov equation. 

The discrete-time Lyapunov equation is extremely well-studied in mathematics and engineering, where it is used for control of macroscopic objects \cite{Control_theory_book1,Control_theory_book2}. Surprisingly we encounter the same equation in our fundamental description of microscopic quantum objects (see also \cite{Archak_2022,Camasca_2021}). The NESS can be obtained directly by solving the discrete-time Lyapunov equation, for which efficient algorithms are already available in standard scientific computing languages like python, mathematica, matlab etc.  With increase in $\tau$, the NESS from PReB can be shown to converge to that from the continuous-time autonomous process. This gives an alternate efficient way to obtain NESS results even in such cases. Further, the discrete time Lyapunov description also points to connections with the standard NEGF description of the continuous-time autonomous process. 
\\

\noindent
{\it Further works ---}Our results open a wide range of directions for further works. As already alluded to, the role of anti-Zeno physics, and the effect of having finite-size baths are interesting questions that require further investigations. Another fundamental point is the role of fluctuations. In the continuous-time  autonomous limit, increasing the efficiency of a heat engine at finite power can be shown to increase fluctuations in power \cite{Pietzonka_2018,Giacomo_2019,Timpanaro_2019}. However, to our knowledge, such a result is not available for cyclic heat engines at arbitrary speed of driving. Therefore, whether a similar trade-off with fluctuations hold for the QTMs at the NESS of the PReB process is unclear at present. Furthermore, in the heat engine regime, it will be interesting to explore how to exploit the power extracted by coupling to a load \cite{VanHorne_2020,Watanabe_2020}, for example, to charge a battery \cite{Barra_2019,Hovhannisyan_2020,Bhattacharjee_2021,BatMark}. In the refrigerating regime, it will be insightful to investigate the effect that non-Markovian dynamics plays in the cooling rate~\cite{Huber_Cooling}. Investigations in these directions will be carried out in future works.
\\

\noindent
{\it Acknowledgements ---} A.P thanks Gabriele de Chiara, Michael Zwolak, Marek Rams and Gabriela Wojtowicz for useful discussions. A.P acknowledges funding from the European Union's Horizon 2020 research and innovation programme under the Marie Sklodowska-Curie Grant Agreement No. 890884. A.P also acknowledges funding from the Danish National Research Foundation through the Center of Excellence ``CCQ'' (Grant agreement no.: DNRF156). S.C. gratefully acknowledges the Science Foundation Ireland Starting Investigator Research Grant “SpeedDemon” (No. 18/SIRG/5508) for financial support. J.P is grateful for financial support from Grant PGC2018-097328-B-100 funded by MCIN/AEI/ 10.13039/501100011033 and, as appropriate, by “ERDF A way of making Europe”, by the “European Union”. G. G. acknowledges fundings from FQXi and DFG Grant No. FOR2724 and from European Unions Horizon
2020 research and innovation programme under the Marie
Sklodowska-Curie Grant Agreement No. 101026667. J. G. is supported by a SFI-Royal Society University Research Fellowship, by the European Research Council Starting Grant ODYSSEY (Grant Agreement No. 758403) and the EPSRC-SFI joint project QuamNESS.

\section*{Appendix}

\appendix

\section{General microscopic dynamics and thermodynamics of open quantum systems}
\label{Appendix:general_definitions_for_thermodynamics}
From the discussion in Sec.~\ref{Sec: Physically_engineering_PReB}, we see that physically realizing the PReB process requires a set-up governed by a Hamiltonian of the form
$
\hat{H}(t)= \hat{H}_S  + \sum_{\ell}\hat{H}_{SB_\ell}(t) + \sum_{\ell} \hat{H}_{B_\ell},
$
(see Fig.~\ref{fig:schematic}). The following gives definitions of thermodynamic quantities for any general set-up of a system coupled to macroscopic baths governed by such a Hamiltonian. We define the entropy production in the open quantum set-up as \cite{Esposito_2010,Reeb_2014,Landi_2021,Strasberg_tutorial_2021}
\begin{align}
\label{def_entropy_production}
& \Sigma(t) = \lim_{D_B \rightarrow \infty} D(\hat{\rho}_{\rm tot}(t)  || \hat{\rho}(t) \hat{\rho}_B(0)) \nonumber \\
&=\sum_{\ell} \beta_\ell Q_\ell(t) - \delta S(t) \geq 0,  \\
&\delta S(t) = S(0)- S(t),~~ S(t)=-{\rm Tr}(\hat{\rho}(t)\log \hat{\rho}(t)), \nonumber
\end{align}
where $D(\hat{\rho}_1 || \hat{\rho}_2)= {\rm Tr}(\hat{\rho_1}\log\hat{\rho_1})-{\rm Tr}(\hat{\rho}_1\log\hat{\rho}_2)$ is called the relative entropy between two density matrices $\hat{\rho}_1$ and $\hat{\rho}_2$, $\hat{\rho}_B(0)$ is the product of thermal states of all the baths (see Eq.(\ref{initial_state})), $D_B$ is the number of degrees of freedom in each bath  and 
\begin{align}
\label{def_heat}
Q_\ell (t) =& \lim_{D_B \rightarrow \infty}\left( \braket{\hat{H}_{B_\ell}(t)}-\braket{\hat{H}_{B_\ell}(0)} \right) \nonumber \\
&- \mu_\ell \lim_{D_B \rightarrow \infty}\left( \braket{\hat{N}_{B_\ell}(t)}-\braket{\hat{N}_{B_\ell}(0)} \right),
\end{align}
is the heat dissipated into the $\ell$th bath. Since relative entropy is non-negative, the second law of thermodynamics is satisfied.   It is crucial to note the presence of $D_B \rightarrow \infty$ limit. It is only in this limit that the baths have infinite capacity for heat and particles and hence temperatures and the chemical potentials can be considered constant, despite strong system-bath coupling. The importance of the $D_B\rightarrow \infty$ limit in using Eq.(\ref{def_entropy_production}) as the definition of entropy production has been highlighted in previous works \cite{Reeb_2014,Strasberg_2021,Strasberg_tutorial_2021}.  In physical realization of the PReB process, we anyway need the presence of macroscopic thermal baths. So, this limit is naturally satisfied. 

 Defining the energy and particle currents from the baths as
\begin{align}
\label{def_currents}
&\hat{J}_{B_\ell\rightarrow S}(t) = -\frac{d\hat{H}_{B_\ell}}{dt}=i[\hat{H}_{B_\ell},\hat{H}(t)],\nonumber \\
&\hat{I}_{B_\ell\rightarrow S}(t) = -\frac{d\hat{N}_{B_\ell}}{dt}=i[\hat{N}_{B_\ell},\hat{H}(t)] ,\nonumber \\
&J_{B_\ell\rightarrow S}(t)=\lim_{D_B \rightarrow \infty}{\rm Tr}\left(\hat{\rho}_{\rm tot} \hat{J}_{B_\ell\rightarrow S}(t)\right), \nonumber \\
&I_{B_\ell\rightarrow S}(t)=\lim_{D_B \rightarrow \infty}{\rm Tr}\left(\hat{\rho}_{\rm tot} \hat{I}_{B_\ell\rightarrow S}(t)\right),
\end{align}
the heat dissipated into the $\ell$th bath can be written as
\begin{align}
\label{def_heat2}
Q_\ell(t) = -\int_0^t ds \Big(J_{B_\ell\rightarrow S}(s)- \mu_\ell I_{B_\ell\rightarrow S}(s)\Big).
\end{align}
The chemical work done on the system is given by
\begin{align}
\label{def_chemical_work}
W_{\rm chem}(t) &= -\lim_{D_B \rightarrow \infty}\sum_\ell \mu_\ell \left(\braket{ \hat{N}_{B_\ell}(t)}-\braket{ \hat{N}_{B_\ell}(0)} \right)\nonumber \\
&= \sum_\ell \mu_\ell \int_0^t ds I_{B_\ell\rightarrow S}(s).
\end{align}
The work done by external forces on the system due to control of system-bath coupling is
\begin{align}
\label{def_external_work}
W_{\rm ext}(t) &= \lim_{D_B \rightarrow \infty}\left(\braket{ \hat{H}(t)} - \braket{ \hat{H}(0)}\right)\nonumber \\
&=\lim_{D_B \rightarrow \infty}\int_0^{t} ds \frac{d\braket{\hat{H}(s)}}{ds} \nonumber \\
&=\lim_{D_B \rightarrow \infty}\int_0^t ds \sum_{\ell}{\rm Tr}\left(\frac{\partial\hat{H}_{SB_\ell}(s)}{\partial s} \hat{\rho}_{\rm tot} \right).
\end{align}
The total work is
\begin{align}
\label{def_total_work}
W(t)=W_{\rm ext}(t) + W_{\rm chem}(t).
\end{align}
The change in internal energy is 
\begin{align}
\label{def_internal_energy}
\delta U(t) = &\lim_{D_B \rightarrow \infty}\Big[\braket{\hat{H}_{S}(t)}- \braket{\hat{H}_{S}(0)}\nonumber \\
&+ \sum_{\ell} \left(\braket{\hat{H}_{SB_\ell}(t)}-\braket{\hat{H}_{SB_\ell}(0)}\right)\Big].
\end{align}
With these definitions it can be checked that the first law of thermodynamics is satisfied,
\begin{align}
\label{first_law}
\delta U(t) = W(t)-Q(t),~~Q(t)=\sum_{\ell} Q_\ell(t).
\end{align}
We will like to mention that all quantities involved in the above definitions of work, heat, entropy production can be obtained in terms of expectation values of system operators and the currents from the baths. These quantities can be calculated without knowing all microscopic details of the baths, but rather, knowing only some macroscopic properties, like the initial temperatures and chemical potentials of the baths and the bath spectral functions. Further, these definitions do not assume weak system-bath coupling, or Markovian dynamics of the system, or any particular slowness or fastness of the time-dependence. Specifically, as shown in Appendix~\ref{Appendix:PReB_thermo}, for the kind of time-dependence in the PReB process, all the above thermodynamic quantities at any given time can be calculated by recursively using one copy of the finite-size environments for each bath.  

\section{Quantum thermodynamics of the PReB process}
\label{Appendix:PReB_thermo}

We apply the formalism in Appendix~\ref{Appendix:general_definitions_for_thermodynamics}, to the physical description of the PReB process described in Sec.~\ref{Sec: Physically_engineering_PReB}. Work done on the system, heat dissipated into the baths, change in internal energy and entropy production can be after $n$ steps can be written as sum of the same for the each step,
\begin{align}
\label{total_thermodynamic_quantities}
& \delta U(n\tau) =\sum_{m=1}^n U^{(m)},~W_{\rm ext}(n\tau)  = \sum_{m=1}^n W_{\rm ext}^{(m)} \nonumber \\
& W_{\rm chem}(n\tau) = \sum_{m=1}^n W_{\rm chem}^{(m)},~Q_\ell (n \tau) = \sum_{m=1}^n Q_\ell^{(m)}, \nonumber \\
& W_{\rm ext}(n\tau)  = \sum_{m=1}^n W_{\rm ext}^{(m)},~ \delta S(n\tau) = \sum_{m=1}^n \delta S^{(m)}, \nonumber \\
& \Sigma(n\tau) = \sum_{m=1}^n  \Sigma^{(m)},
\end{align}
where $U^{(m)}$, $W_{\rm ext}^{(m)}$, $Q_\ell^{(m)}$, $W_{\rm chem}^{(m)}$,   $\delta S^{(m)}$, $\Sigma^{(m)}$ are the change in internal energy, work done by the external forces, heat dissipated into the $\ell$th bath, the chemical work done on the system, the change in von-Neumann entropy of the system and the entropy production of the system respectively in the $m$th step. These quantities, for the PReB process are given by
\begin{align}
& \delta U^{(m)}=\braket{\hat{H}_{S}(m\tau)}- \braket{\hat{H}_{S}\big((m-1)\tau\big)}, \nonumber \\
& W_{\rm ext}^{(m)}=\Big[\braket{ \hat{H}_S(m\tau)} - \braket{ \hat{H}_S\big( (m-1)\tau \big)} \nonumber \\
& \qquad + \sum_\ell \left(\braket{ \hat{H}_{B_\ell}(m\tau)} - \braket{ \hat{H}_{B_\ell}\big( (m-1)\tau\big)} \right) \Big], \nonumber \\
& W_{\rm chem}^{(m)}=\nonumber \\
&\qquad - \sum_\ell \mu_\ell \left( \braket{\hat{N}_{B_\ell}(m\tau)}-\braket{\hat{N}_{B_\ell}\big((m-1)\tau \big)} \right) \nonumber \\
& Q_\ell^{(m)}=\left( \braket{\hat{H}_{B_\ell}(m\tau)}-\braket{\hat{H}_{B_\ell}\big((m-1)\tau \big)} \right)\nonumber \\
&\qquad - \mu_\ell \left( \braket{\hat{N}_{B_\ell}(m\tau)}-\braket{\hat{N}_{B_\ell}\big((m-1)\tau \big)} \right), \nonumber\\
& S^{(m)} = S(m\tau) - S\big( (m-1)\tau\big), \nonumber\\
& \Sigma^{(m)} = \sum_\ell \beta_\ell Q_\ell^{(m)} - \delta S^{(m)}.
\end{align}
In obtaining above, all expectation values required for the thermodynamic quantities are calculated just after switching off system-bath coupling for previous step and just before switching on system-bath coupling for the next step. So, the above expressions of change in internal energy and external work do not explicitly involve the system-bath coupling Hamiltonian. However, using conservation of energy, the external work can be recast in terms of only the system-bath coupling Hamiltonian. To see this, note that during the $m$th step of the process, there is no explicit time dependence in the Hamiltonian and so the total energy of the full set-up is conserved,
\begin{align}
&\braket{ \hat{H}_S(m\tau)}+\sum_{\ell} \Big[ \braket{ \hat{H}_{B_\ell}(m\tau)}+\braket{ \hat{H}_{SB_\ell}(m\tau)} \Big] \nonumber \\
& = \braket{ \hat{H}_S\big((m-1)\tau\big)}+\sum_{\ell} \Big[ \braket{ \hat{H}_{B_\ell}\big((m-1)\tau\big)} \nonumber \\
& \hspace*{15pt}+\braket{ \hat{H}_{SB_\ell}\big((m-1)\tau\big)} \Big].
\end{align}
Using this, the expression for external work done in the $m$th step of the PReB process can be simplified to
\begin{align}
& W_{\rm ext}^{(m)}=\nonumber \\
& \qquad-\sum_\ell \Big[\braket{ \hat{H}_{SB_\ell}(m\tau)} - \braket{ \hat{H}_{SB_\ell}\big( (m-1)\tau \big)} \Big].
\end{align}

During the $m$th step, the only change in energy and number of particles in the baths occur in the finite-size chains attached to the system during that step because the corresponding residual baths are left unaffected during the time interval $\tau$ (see Fig.~\ref{fig:schematic}). Further, the only time-dependence in the Hamiltonian during this step is the switching on and off of the coupling to the  corresponding finite-size chains.  Using the above facts, it is possible to write the thermodynamic quantities for the $m$th step in terms of quantities calculated by using one copy of the finite-size chains for each bath. To this end, we define the corresponding Hamiltonian
\begin{align}
\label{H_m}
\hat{H}_m = \hat{H}_S + \sum_{\ell}  \hat{H}_{B_{\ell m}} + \sum_\ell \hat{H}_{SB_{\ell m}},
\end{align}
where $\hat{H}_{B_{\ell m}}$ is the Hamiltonian of the finite-size chain of $\ell$th bath that affects the system during the $m$th step, and  $\hat{H}_{SB_{\ell m}}$ is the corresponding system-bath coupling.
The $m$th step of PReB is then given by
\begin{align}
\label{single_step_of_PReB}
& \hat{\rho}^{(m)}={\rm Tr_B} \left(\hat{\rho}_{\rm tot}^{(m)}\right)=\hat{\Lambda}(\tau)[\hat{\rho}^{m-1}],\nonumber \\
&\hat{\rho}_{\rm tot}^{(m)}(\tau) \nonumber \\
&= e^{-i\hat{H}_m\tau} \hat{\rho}^{(m-1)}\prod_\ell\frac{e^{-\beta_\ell(\hat{H}_{B_{\ell m}}-\mu_\ell \hat{N}_{B_{\ell m}})}}{Z_{B_{\ell m}}} e^{i\hat{H}_m\tau},
\end{align}
where $\hat{N}_{B_{\ell m}}$ is the total number operator for the $\ell$th finite-size chain affecting the system during the $m$th step and ${\rm Tr_B}(\ldots)$ now refers to trace over the finite-sized chains. Here $m=1,2,3,\ldots$, and $\hat{\rho}^{0} = \hat{\rho}(0)$.  Now, the thermodynamic quantities for the $m$th step can be written as
\begin{align}
\label{single_step_defs}
&\delta U^{(m)}=\hspace*{-5pt} {\rm Tr}\left(\hat{H}_S \hat{\rho}_{\rm tot}^{(m)}(\tau)\right)-{\rm Tr}\left(\hat{H}_S \hat{\rho}_{\rm tot}^{(m)}(0)\right), \nonumber \\
& \delta S^{(m)} = S^{(m-1)}-S^{(m)}, \nonumber \\
&S^{(m)}=-{\rm Tr}\left(\hat{\rho}^{(m)}\log \hat{\rho}^{(m)}\right), \nonumber \\
& W^{(m)}_{\rm ext}=\nonumber\\
&-\sum_{\ell} \Big[{\rm Tr}\left(\hat{H}_{SB_{\ell m}} \hat{\rho}_{\rm tot}^{(m)}(\tau)\right)-{\rm Tr}\left(\hat{H}_{SB_{\ell m}} \hat{\rho}_{\rm tot}^{(m)}(0)\right)\Big],\nonumber \\
& W^{(m)}_{\rm chem}=\nonumber\\
&-\sum_{\ell}\mu_\ell\Big[ {\rm Tr}\left(\hat{N}_{B_{\ell m}} \hat{\rho}_{\rm tot}^{(m)}(\tau)\right)-{\rm Tr}\left(\hat{N}_{B_{\ell m}} \hat{\rho}_{\rm tot}^{(m)}(0)\right) \Big], \nonumber \\
&Q_{\ell}^{(m)}=\Big[ {\rm Tr}\left(\hat{H}_{B_{\ell m}} \hat{\rho}_{\rm tot}^{(m)}(\tau)\right)-{\rm Tr}\left(\hat{H}_{B_{\ell m}} \hat{\rho}_{\rm tot}^{(m)}(0)\right) \Big] \nonumber \\
& -\mu_\ell\Big[ {\rm Tr}\left(\hat{N}_{B_{\ell m}} \hat{\rho}_{\rm tot}^{(m)}(\tau)\right)-{\rm Tr}\left(\hat{N}_{B_{\ell m}} \hat{\rho}_{\rm tot}^{(m)}(0)\right) \Big], \nonumber \\
& \Sigma^{(m)}=\sum_\ell \beta_\ell Q_{\ell}^{(m)} - \delta S^{(m)}. 
\end{align}
The above results show that, due to the particular nature of the PReB process, the thermodynamic quantities can be calculated by recursively using only one copy of the finite-size chain corresponding to each bath.

Note from Eq.(\ref{single_step_of_PReB}) that if we take $m=1$ and $\tau=t$, we describe the continuous-time autonomous process in Eq.(\ref{def_Lambda}). With this substitution, the above expressions for the thermodynamic quantities carry over to the continuous-time autonomous process (provided the system-bath coupling is switched off at time $t$).

The NESS of the PReB process is obtained in the limit of $n\rightarrow \infty$ for a finite value of $\tau$. The NESS satisfies Eq.(\ref{def_PReB_NESS}). Consequently, at NESS, in a single step, the change in internal energy and the change in entropy of the system are zero, while work done on the system and the heat dissipated into the baths are non-zero constants. As a result, the work done on the system and the heat dissipated into the baths diverge as $n$. Thus we have
\begin{align}
\label{PReB_NESS_thermodynamic_defs_appendix}
& P_{\rm ext}=\lim_{n\rightarrow\infty}\frac{W_{\rm ext}(n\tau)}{n\tau},P_{\rm chem}=\lim_{n\rightarrow\infty}\frac{W_{\rm chem}(n\tau)}{n\tau},\nonumber \\
& P = P_{\rm ext}+P_{\rm chem}, \nonumber \\
&\dot{Q}_{\ell}=\lim_{n\rightarrow\infty}\frac{Q_{\rm \ell}(n\tau)}{n\tau},~~\sigma=\lim_{n\rightarrow\infty}\frac{\Sigma(n\tau)}{n\tau}, \nonumber \\
&\lim_{n\rightarrow\infty}\frac{ \delta U(n\tau)}{n\tau} = 0,~~\lim_{n\rightarrow\infty} \frac{\delta S(n\tau)}{n\tau} = 0.  
\end{align}
In above,  $P_{\rm ext}$ is the power input into the system due to external switching on and off of the system-bath couplings at NESS, $P_{\rm chem}$ is the power input due to the chemical potentials of the baths at NESS, $\dot{Q}_{\ell}$ is the heat dissipation rate at NESS, $\sigma$ is the entropy production rate at NESS. In terms of these rates, the first law of thermodynamics becomes
\begin{align}
\label{first_law_NESS_appendix}
& P=\sum_\ell \dot{Q}_{\ell},
\end{align}
while the second law gives
\begin{align}
\label{second_law_NESS_appendix}
\sigma=\sum_{\ell} \beta_\ell \dot{Q}_{\ell} \geq 0. 
\end{align}
These are the standard expressions for the first law and the second law of thermodynamics at NESS. These expressions are valid both for at the NESS of the PReB process and at the NESS of the continuous-time autonomous process corresponding to $n=1$, $\tau\rightarrow \infty$. Utilizing the results in Eqs.(\ref{total_thermodynamic_quantities}) and (\ref{single_step_defs}), the expression for power and heat currents given in Eq.(\ref{PReB_NESS_thermodynamics}) can be obtained.  

\section{Dynamics and NESS of PReB process for Gaussian systems}
\label{Appendix: Gaussian_PReB} 
In this Appendix, we give the derivation of Eq. Eq.(\ref{Gaussian_PReB_system}).
\subsection{The correlation matrix formalism for Gaussian systems}
\label{Appendix:Gaussian_systems}

We consider set-ups, where the Hamiltonian governing the dynamics of the system during the $m$th step is of the form,
\begin{align}
\label{full_set_up_non_interacting_appendix}
\hat{H}_m=\sum_{p,q=1}^{L_S+2L_B} \mathbf{H}_{p,q} \hat{d}_p^\dagger \hat{d}_q,
\end{align}
where we have assumed that there are two environments, each of size $L_B$, attached to the system, and $\hat{d}_p$ is the fermionic annihilation operator of either a system or an environment site. Here, $\mathbf{H}$ is a $(L_S+2L_B) \times (L_S+2L_B)$ dimensional real symmetric matrix, often called the single-particle Hamiltonian.  We define the $(L_S+2L_B) \times (L_S+2L_B)$ dimensional equal time correlation matrix, sometimes called the single-particle density matrix, of the set-up,
\begin{align}
\mathbf{C}_{p,q}(t) = {\rm Tr}\left(\hat{\rho}_{\rm tot}(t) \hat{d}_p^\dagger \hat{d}_q \right).
\end{align}
Both the matrix $\mathbf{H}$ and the correlation matrix $\mathbf{C}$ can be written in the following block form,
\begin{align}
&\mathbf{H}=\left[
\begin{array}{ccc}
\mathbf{H}_S & \mathbf{H}_{SB_1} & \mathbf{H}_{SB_2} \\
\mathbf{H}_{SB_1}^T & \mathbf{H}_{B_1} & \mathbf{0} \\
\mathbf{H}_{SB_2}^T & \mathbf{0} & \mathbf{H}_{B_2}
\end{array}
\right], \nonumber \\
&\mathbf{C}=\left[
\begin{array}{ccc}
\mathbf{C}_S & \mathbf{C}_{SB_1} & \mathbf{C}_{SB_2} \\
\mathbf{C}_{SB_1}^\dagger & \mathbf{C}_{B_1} & \mathbf{C}_{B_1 B_2} \\
\mathbf{C}_{SB_2}^\dagger & \mathbf{C}_{B_1 B_2}^\dagger & \mathbf{C}_{B_2}
\end{array}
\right].
\end{align}
Here, the $L_S \times L_B$ matrix $\mathbf{H}_{SB_1}$ ($\mathbf{H}_{SB_2}$) gives the coupling between the system and the first (second) bath, $\mathbf{H}_{SB_1}^T$ ($\mathbf{H}_{SB_2}^T$) is its transpose, $\mathbf{H}_{B_1}$ ($\mathbf{H}_{B_2}$) gives the Hamiltonian of the first (second) bath. Similarly, the $L_S \times L_S$ dimensional matrix $\mathbf{C}_S$ gives the system correlation matrix, the $L_S \times L_B$ dimensional matrix   $\mathbf{C}_{SB_1}$ ($\mathbf{C}_{SB_2}$) gives the correlations between the system and the first (second) bath, $\mathbf{C}_{B_1}$ ($\mathbf{C}_{B_2}$) gives the correlation matrix corresponding to the first (second) bath, $\mathbf{C}_{B_1 B_2}$ gives the correlations between the two baths. The elements of the system correlation matrix are
\begin{align}
\mathbf{C}_{S_{p,q}}(t)={\rm Tr}\left(\hat{\rho}_{\rm tot}(t) \hat{c}_p^\dagger \hat{c}_q \right).
\end{align}
The elements of the other correlation matrices are given likewise. The time evolution of $\mathbf{C}(t)$ can be written in the form,
\begin{align}
\frac{d \mathbf{C}}{dt}=-i[\mathbf{C},\mathbf{H}] \Rightarrow \mathbf{C}(t) = e^{i\mathbf{H} t} \mathbf{C}(0) e^{-i\mathbf{H} t}.
\end{align}
For Gaussian states, given $\mathbf{C}(t)$, the density matrix can be uniquely obtained \cite{Peschel_2001,Dhar_2012,Peschel_2017}.  Further, to obtain the state of any subsystem of the set-up, it is only required to restrict $\mathbf{C}(t)$ to the sites corresponding to the subsystem. 

The thermal states of the baths are Gaussian. If the initial state of the system is also Gaussian then, due to the form of the Hamiltonian, the state of the set-up remains Gaussian throughout the time evolution. In such cases, we can describe the PReB process in terms of only the $(L_S+2L_B) \times (L_S+2L_B)$ dimensional correlation matrix, instead of the $2^{(L_S+2L_B)} \times 2^{(L_S+2L_B)}$ dimensional density matrix $\hat{\rho}_{\rm tot}(t)$. 

\subsection{PReB in terms of the correlation matrix}
\label{Appendix:PReB_from_correlation_matrix}

To describe the dynamics, it is sufficient to consider only one copy of the tight-binding chains representing the baths, which are periodically disconnected and instantaneously refreshed to their original initial thermal states. So, at the end of each step, the set-up is at a product state of the system and the thermal states of the finite-size baths.  Thus, at the end of the $m$th step the correlation matrix of the set-up has to be of the form
\begin{align}
\label{initial_state_PReB_appendix}
\mathbf{C}^{(m)} = \left[
\begin{array}{ccc}
\mathbf{C}_S^{(m)} & \mathbf{0} & \mathbf{0} \\
\mathbf{0} & \mathbf{C}_{B_1}^{\rm therm} & \mathbf{0} \\
\mathbf{0} & \mathbf{0} & \mathbf{C}_{B_2}^{\rm therm}
\end{array}
\right],
\end{align}
where $\mathbf{C}_S^{(m)}$ is the correlation matrix giving the state of the system after $m$ steps, while, $\mathbf{C}_{B_1}^{\rm therm}$ ($\mathbf{C}_{B_2}^{\rm therm}$) is the correlation matrix corresponding to the thermal state of the first (second) bath, 
\begin{align}
&\left(\mathbf{C}_{B_\ell}^{\rm therm}\right)_{p,q} = {\rm Tr}\left(\hat{b}^\dagger_{\ell,p} \hat{b}_{\ell,q} \frac{e^{-\beta_\ell(\hat{H}_{B_{\ell}}-\mu_\ell \hat{N}_{B_{\ell}})}}{Z_{B_{\ell}}}\right), \nonumber \\
& \ell=\{1,2\}.
\end{align}
To take the next step, we have to evolve the correlation matrix up to a time $\tau$, and then retain only the block corresponding to the system while replacing all other blocks by the same elements as in Eq.(\ref{initial_state_PReB_appendix}). It can be checked that this entire operation can be written as the following matrix equation
\begin{align}
\label{Gaussian_PReB_full}
& \mathbf{C}^{(m+1)} = \mathbf{M}^\dagger  \mathbf{C}^{(m)} \mathbf{M} + \mathbf{P},~~{\rm with}~~
 \mathbf{M} = e^{-i\mathbf{H}\tau} \mathbf{A}, \nonumber \\
&\mathbf{A}=\left[
\begin{array}{ccc}
\mathbb{I} & \mathbf{0} & \mathbf{0} \\
\mathbf{0}  & \mathbf{0} & \mathbf{0} \\
\mathbf{0} & \mathbf{0} & \mathbf{0}
\end{array}
\right],
\mathbf{P}=\left[
\begin{array}{ccc}
 \mathbf{0} & \mathbf{0} & \mathbf{0} \\
 \mathbf{0} & \mathbf{C}_{B_1}^{\rm therm} & \mathbf{0} \\
\mathbf{0} & \mathbf{0} & \mathbf{C}_{B_2}^{\rm therm}
\end{array}
\right],
\end{align}
where $\mathbb{I}$ is the $L_S$ dimensional Identity matrix.  Now, carrying out explicitly the operation in Eq.(\ref{Gaussian_PReB_full}), the correlation matrix giving the state of the system after $m+1$ steps can be seen to be given by Eq.(\ref{Gaussian_PReB_system}).

\section{Heat engines and refrigerators with two baths}
\label{Appendix:heat_engines_and_refrigerators}

Here we discuss how the Carnot bound on efficiency (coefficient of performance) of a heat engine (refrigerator) with two baths arises out of the general expression for first law and second law of thermodynamics at NESS. For this we will assume that the first bath is at a higher temperature and the second bath is at a lower temperature,
\begin{align}
\label{hot_and_cold_appendix}
\beta_2=\beta_1 + \Delta \beta,~~\Delta \beta = \beta_2 - \beta_1 > 0.
\end{align} 
Further, we reiterate our sign convention: work done on the system is positive, whereas, work done by the system is negative; heat dissipated into bath is positive, while heat extracted from bath is negative.

\subsection{Heat engine}
The heat engine regime corresponds to the case where work is done by the system, thereby the steady state power being negative,
\begin{align}
P=P_{\rm ext}+P_{\rm chem}<0.
\end{align}
Combining the first law and the second law for thermodynamics at NESS using Eq.(\ref{hot_and_cold_appendix}), we write the entropy production rate as
\begin{align}
\label{entropy_production_heat_engine}
\sigma = \beta_2 P - \Delta \beta \dot{Q}_{1} \geq 0.
\end{align}
 This shows that power cannot be extracted without any temperature bias, and for heat engine, heat must flow from the hot bath into the system for the heat engine $\dot{Q}_{1} <0$. The efficiency $\eta$ of the heat engine is then the ratio of extracted power over the rate of heat flow from hot bath. From Eq.(\ref{entropy_production_heat_engine}), it is easy to see that the efficiency is bounded from above by the Carnot efficiency $\eta_c$,
\begin{align}
\eta = \frac{P}{\dot{Q}_1},~~\eta_c = 1-\frac{\beta_1}{\beta_2},~~\eta \leq \eta_c.
\end{align}

\subsection{Refrigerator}
The refrigerating regime corresponds to the case where the heat flows out of the cold bath, thereby cooling it further. This means
\begin{align}
\dot{Q}_{2} <0.
\end{align}
As before, we can combine the first law and the second law of thermodynamics at NESS using Eq.~(\ref{hot_and_cold_appendix}) and write the entropy production rate as 
\begin{align}
\label{entropy_production_refrigerator}
\sigma = \beta_1 P + \Delta \beta \dot{Q}_{2} \geq 0.
\end{align}
This dictates that
$
P=P_{\rm ext}+P_{\rm chem} > 0.
$
Thus, power needs to be input to extract heat from the cold bath in order for the refrigerator to work. The coefficient of performance ($COP$) of the refrigerator is the ratio of cooling rate over input power. From Eq.~(\ref{entropy_production_refrigerator}), it is easy to see that the $COP$ is bounded from above,
\begin{align}
&COP = \frac{-\dot{Q}_{2}}{P},~~COP \leq COP_c, \nonumber \\
& COP_c = \frac{1}{\beta_2/\beta_1-1},
\end{align}
where $COP_c$ is the coefficient of performance of the Carnot refrigerator.

\bibliographystyle{apsrev4-1}
\bibliography{ref_PReB2}
\end{document}